\documentclass[]{aastex631}

\shorttitle{Grid-Based Retrievals}
\shortauthors{Susemiehl et al.}

\graphicspath{{./}{figures/}}

\usepackage{algorithm}
\usepackage{float}
\usepackage[normalem]{ulem}

\begin{document}

\title{Grid-Based Atmospheric Retrievals for Reflected-Light Spectra of Exoplanets using PSGnest}

\author{Nicholas Susemiehl}
\affiliation{Southeastern Universities Research Association, Washington, DC, USA.}
\affiliation{Center for Research and Exploration in Space Science and Technology, NASA Goddard Space Flight Center, Greenbelt, MD, USA.}
\affiliation{NASA Goddard Space Flight Center, Greenbelt, MD, USA.}

\author[0000-0002-8119-3355]{Avi M. Mandell}
\affiliation{NASA Goddard Space Flight Center, Greenbelt, MD, USA.}

\author[0000-0002-2662-5776]{Geronimo L. Villanueva}
\affiliation{NASA Goddard Space Flight Center, Greenbelt, MD, USA.}

\author{Giuliano Liuzzi}
\affiliation{NASA Goddard Space Flight Center, Greenbelt, MD, USA.}
\affiliation{Dep. of Physics, American University, Washington, DC, USA.}

\author{Michael Moore}
\affiliation{NASA Goddard Space Flight Center, Greenbelt, MD, USA.}
\affiliation{Business Integra, Inc., Bethesda, MD, USA.}

\author{Tyler Baines}
\affiliation{Southeastern Universities Research Association, Washington, DC, USA.}
\affiliation{Center for Research and Exploration in Space Science and Technology, NASA Goddard Space Flight Center, Greenbelt, MD, USA.}
\affiliation{NASA Goddard Space Flight Center, Greenbelt, MD, USA.}

\author{Michael D. Himes}
\affiliation{Planetary Sciences Group, Department of Physics, University of Central Florida, Orlando, FL, USA.}
\affiliation{NASA Postdoctoral Program Fellow, NASA Goddard Space Flight Center, Greenbelt, MD, USA.}

\author{Adam J. R. W. Smith}
\affiliation{Southeastern Universities Research Association, Washington, DC, USA.}
\affiliation{Center for Research and Exploration in Space Science and Technology, NASA Goddard Space Flight Center, Greenbelt, MD, USA.}
\affiliation{NASA Goddard Space Flight Center, Greenbelt, MD, USA.}

\begin{abstract}

Techniques to retrieve the atmospheric properties of exoplanets via direct observation of their reflected light have often been limited in scope due to computational constraints imposed by the forward-model calculations. We have developed a new set of techniques which significantly decreases the time required to perform a retrieval while maintaining accurate results. We constructed a grid of 1.4 million pre-computed geometric albedo spectra valued at discrete sets of parameter points. Spectra from this grid are used to produce models for a fast and efficient nested sampling routine called PSGnest. Beyond the upfront time to construct a spectral grid, the amount of time to complete a full retrieval using PSGnest is on the order of seconds to minutes using a personal computer. An extensive evaluation of the error induced from interpolating intermediate spectra from the grid indicates that this bias is insignificant compared to other retrieval error sources, with an average coefficient of determination between interpolated and true spectra of 0.998. We apply these new retrieval techniques to help constrain the optimal bandpass centers for retrieving various atmospheric and bulk parameters from a LuvEx-type mission observing several planetary archetypes. We show that spectral observations made using a 20\% bandpass centered at 0.73 microns can be used alongside our new techniques to make detections of $H_2O$ and $O_2$ without the need to increase observing time beyond what is necessary for a signal-to-noise ratio of 10. The methods introduced here will enable robust studies of the capabilities of future observatories to characterize exoplanets.

\end{abstract}


\section{Introduction} \label{sec:intro}

Spectral or atmospheric retrieval is one of the most direct and powerful methods available for remotely exploring the composition of the atmospheres and surfaces of extrasolar planets. The objective of these retrievals is to disentangle the spectral signatures of atmospheric and surface constituents, as well as atmospheric parameters such as temperature and pressure. Doing so enables the constraint of the planet's atmospheric composition and structure as well as its bulk planetary parameters \citep{2018haex.bookE.104M}. 
Spectral retrieval methodologies developed for characterizing exoplanets have been adapted from existing and highly effective tools used for Solar System studies \citep{https://doi.org/10.1029/RG014i004p00609}. The first works to retrieve the atmospheres of exoplanets used optimal estimation schemes based on Solar System retrievals \citep{2009ApJ...707...24M}, but later works showed Bayesian frameworks to be more successful for the highly unconstrained and degenerate planetary parameters characteristic of exoplanet science \citep{2012ApJ...753..100B}. These methods have been used to constrain temperatures and abundances of gas giant atmospheres \citep[e.g.,][]{2014ApJ...793...33L, 2014Sci...346..838S, 2015ApJ...814...66K, 2017ApJ...847L...3O, 2018ApJ...855L..30A, 2018ApJ...866...27L, 2019AJ....157..114B, 2020MNRAS.497.5155W,  2022PSJ.....3...80H, HimesHarrington2022apjWASP12b}, model atmospheric winds \citep{2020A&A...633A..86S}, and detect water in the atmosphere of a hot Jupiter \citep{2020MNRAS.497.5155W}. 

Due to the immediate and burgeoning pool of spectroscopic data for transiting planets, the majority of exoplanet spectral retrieval tools have been developed to tackle data for the physical conditions and radiative transfer geometry probed by transiting planet measurements. This has enabled the characterization of a variety of planets with high planet-star radius ratios relative to those probed by other detection techniques. While the transit method has been successful in detecting nearly 4000 exoplanets to date (almost 80\% of all known exoplanets) \footnote{https://exoplanetarchive.ipac.caltech.edu/}, its limitations (combined with those of other prevalent detection techniques such as radial velocity) have created a sizable gap in the mass-period discovery space. To date, no Earth-sized or even Neptune-sized planets have been detected in the habitable zones of Sun-like stars \citep{2021AJ....161..150C} due to the extreme contrast between the parent star and the planet (of order $10^{-10}$ contrast \citet{2021AJ....161..150C}). However, with the advent of new instruments for suppressing light from the central star, direct imaging observations of reflected planetary light have begun to yield results. Current instruments such as the Very Large Telescope's Spectro-Polarimetric High-Contrast Exoplanet Research \citep{2019A&A...631A.155B} and the Gemini South Telescope's Gemini Planet Imager \citep{2015Sci...350...64M} have been able to examine dozens of brown dwarfs and distant giant planets using direct imaging, and the Nancy Grace Roman Space Telescope \citep[formerly WFIRST;][]{2013arXiv1305.5425S} is expected to push this down to Jupiter analogs. Future missions such as the LUVOIR and HabEx concepts studied as part of the recent Astro2020 Decadal Survey \citep{2021pdaa.book.....N} will have the ultimate goal of charting a path to the detection and characterization of Earth-like planets orbiting Sun-like stars in reflected light. Furthermore, using these instruments to constrain the abundance of biosignatures such as $O_2$ and $O_3$ will help us explore the possibility of extraterrestrial life in the universe \citep{2018AsBio..18..663S}. Now that the first detections of an Earth-like planet orbiting a Sun-like star could be made in the near future, it is necessary to develop and validate retrieval techniques capable of accurately quantifying the atmospheric compositions and bulk properties of these planets \citep{2022arXiv220413816D}.   \par

Retrieval studies are also useful for evaluating future mission yields. Such studies have been performed to assess the science return for giant gaseous exoplanets in reflected light \citep{2016AJ....152..217L, 2017PASP..129c4401N}, examine how the constraints yielded by retrievals of Earth-like exoplanets vary for different noise and resolving power levels \citep{2018AJ....155..200F}, investigate potential yields for atmospheric water constraints \citep{2020AJ....159...36S}, and validate retrieval methods using Solar System analogs \citep{https://doi.org/10.48550/arxiv.2204.04231}. However, one of the major challenges for evaluating future mission yields with atmospheric retrieval studies is the computational runtime required to effectively examine a wide range of instrument and observing scenarios. Bayesian retrievals compare the observed spectrum to thousands or millions of model spectra using a likelihood function. These model spectra are often simulated using complex radiative transfer codes in real-time. A typical cloudy spectrum can take about a minute to generate using state-of-the-art radiative transfer tools \citep[e.g.,][]{2018JQSRT.217...86V}. Modeling scattering processes, an important contribution for reflected-light spectra at short wavelengths, increases the computational cost further. If 10,000 models are required to perform a retrieval, real-time model generation can cause the retrieval to take a week. Other works have reported similarly long retrieval runtimes \citep{2018AJ....155..200F}. In this regime, performing the variety of retrievals necessary to explore multiple instrument designs and wavelength ranges becomes untenable.

Several recent studies have investigated various means to accelerate retrievals. Machine learning methods are becoming more prevalent throughout the field and have frequently been applied to retrieval studies \citep[e.g.][]{2016ApJ...820..107W, 2018AJ....156..268Z, 2018NatAs...2..719M, 2018arXiv181103390S, 2019AJ....158...33C, 2020AJ....159..192F}. While these methods can reduce compute times by several orders of magnitude, this often comes at the cost of the accuracy of the resulting posterior distributions. Attempts to remedy this include \cite{2022PSJ.....3...91H} which presents retrievals utilizing a traditional Bayesian framework but with the radiative transfer forward model replaced by a neural network. Though slower than other machine learning approaches to retrieval, their method more closely agrees with traditional retrieval methods while still reducing computational costs by orders of magnitude. Other recent works have incorporated variational inference and normalizing flow-based neural networks \citep{2022arXiv220507037H, 2023A&A...672A.147V} which have been shown to produce comparable posterior distributions to more traditional Bayesian methods while significantly accelerating retrievals. On the other hand, an example of an accelerated retrieval tool which does not implement machine learning methods is {\tt rfast} \citep{2022arXiv220404231R}. The {\tt rfast} framework enables the retrieval of a variety of scenarios including reflected light, thermal emission, and transmission observations. {\tt rfast} takes advantage of linear algebra techniques to vectorize most computations, greatly accelerating the time needed to generate a spectrum. {\tt rfast} was validated using Solar System analog data and performs as accurately as radiate transfer-based methods. Direct comparisons between the total runtimes and forward model accuracy of different retrieval methods are difficult to make due to potentially significant differences in model parameterizations and computer hardware, but future inter-comparisons could help to explore these questions.

Another option to accelerate exoplanet atmospheric retrievals which we adopt in this work is to utilize a grid-based approach. Instead of producing spectra in real time, a grid of spectra is pre-generated at defined parameter values. Intermediate spectra are interpolated from this grid for use as the model spectra during the retrieval runtime. These calculations are performed using a linear interpolation scheme which interpolates intermediate parameter values and combines the resulting spectra proportionally. This process induces an interpolation error which is proportional to the distance the interpolated points are from the closest grid points in the multidimensional space. For this reason, it is important to strategically place a sufficient number of grid points for each parameter to minimize this interpolation error. However, each new grid point adds a significant number of spectra to the grid (the total number of spectra in the full grid is equal to the product of the number of grid points for each parameter). Therefore, it is important to carefully choose the number and placement of grid points while also not adding an excessively large number of points due to computational constraints during grid generation. 

A number of studies have produced spectral model grids across a range of planetary parameters \citep{2010DPS....42.4401F, 2017PASP..129d4402K, 2018MNRAS.474.5158G, 10.1093/mnras/sty3001, 2019MNRAS.482.4503G, 2020MNRAS.498.4680G, 2020AJ....160..204S}, and several have incorporated grid-based methodologies into retrieval studies \citep[e.g.][]{2016ApJ...820..107W, 2018NatAs...2..214D, https://doi.org/10.48550/arxiv.2206.12194}. Most commonly, grids are constructed using an equal number of linearly spaced points for each parameter \citep[e.g.][]{2001ApJ...556..357A, 2019MNRAS.482.4503G, 2021jwst.prop.1977M}. Constructing grids in this manner enables the incremental study of spectra as particular parameters are changed in defined steps. However, this approach may not be optimal for retrieval studies which seek to maintain the accuracy of retrievals compared with full radiative transfer calculations by minimizing the interpolation error. Placing grid points evenly ignores the nonlinear effect that changing certain parameters has on the morphology of a spectrum. This method could also under- or over-sample certain regions of the parameter space, resulting in interpolation errors that are either excessively high or grid sizes which are computationally challenging or even infeasible to generate. For these reasons, exoplanet spectral retrieval studies using a grid-based approach may benefit from a different means of choosing the placement of grid points. 

In this work, we present a novel, nonlinear approach to grid construction. We developed an algorithm which iteratively adds additional grid points at the location with the highest interpolation error. This results in the maximum reduction of interpolation error at each step by optimizing the trade-off between computational complexity and interpolation accuracy. While the grid-based techniques we present in this work induce some interpolation error and require an initial computation investment, they reduce total retrieval runtimes to minutes or even seconds (using a standard high-end laptop.). This significant speed-up enables a host of new studies which involve many retrieval runs and will greatly enhance our capabilities to examine the sensitivity of model parameter inference to the expected performance capabilities of near-future observatory missions. Our methods for constructing, evaluating, and deploying this grid are the primary topics of this work. \par

This work is structured as follows: in Section~\ref{grid}, we describe our methods for parameterizing and constructing the grid. In Section~\ref{retrievals}, we present our implementation and evaluation of grid-based retrievals. Section~\ref{bp} describes an application of these methods to a scientific case. We discuss caveats associated with this work in Section~\ref{discussion} and conclude in Section~\ref{conclusions}

\section{Input Spectral Grids} \label{grid}
The goal for our project was to construct a grid of model reflectance spectra at the visible wavelengths, focusing on rocky planets and using a simplified set of planetary parameters; the ranges of parameter values would be centered around an Earth-like case. This grid would allow us to examine the effectiveness of retrieving planetary parameters for the case of a potentially Earth-like planet assuming different observing scenarios.

In order to build a suitable spectral grid, a complete radiative transfer code capable of including all the surface and atmospheric components of interest was required. To this end, we employed the radiative transfer capabilities of the Planetary Spectrum Generator \citep[PSG;][]{2018JQSRT.217...86V,2022fpsg.book.....V} to generate the grid of spectra. PSG is a state-of-the-art radiative transfer suite, incorporating a variety of different spectroscopic methods, opacity databases, and continuum and scattering processes (e.g., Rayleigh, CIAs) to synthesize spectra at different viewing geometries. At the core of the planetary radiative-transfer module of PSG, the PSGDORT module performs multiple-scattering calculations in a layer-by-layer framework. Many spectral databases are available in PSG, but for this study we have employed the molecular parameters from the latest HITRAN-2020 database \citep{hitran20} that are integrated using a correlated-k method. The molecular databases are complemented in the UV/optical with cross-sections from the Max Planck Institute of Chemistry database \citep{Kellerrudek13}. Besides the collision-induced absorption (CIA) bands available in the HITRAN database, the MT\_CKD water continuum is characterized as H$_2$O–H$_2$O and H$_2$O–N$_2$ CIAs \citep{kofmanvilla21}.

\subsection{Grid Parameterization}
\label{param_sec}
The grid of planetary reflectance spectra were computed as geometric albedo spectra (I/F) over a broad bandpass from 0.4 - 1.0 $\mu m$; this wavelength range was chosen to encompass the defined spectral range for the visible channels of the exoplanet imaging instruments from the LUVOIR and HabEx mission studies  We chose a native spectral resolution (R) of 500, which allows grid users to down-sample the spectra to any lower R. The atmospheric layering and vertical temperature and abundance profiles were generated following the methods described in \cite{2020AJ....160..204S}. The atmospheres of these spectra used constant volume mixing ratio (VMR) profiles of $H_2O$, $O_3$, and $O_2$ with an $N_2$ background (i.e. $N_2=1-H_2O-O_3-O_2$), and the temperature profiles are constant at 250 K. The model-top pressure is set to $10^{-4}$ bars for each spectrum. Both clear and ``cloudy" versions of each spectra were created. The cloudy spectra contain isotropic, wavelength independent clouds with a mass mixing ratio of 0.23509 ppm and a particle size distribution peaked at 1 $\mu$m and an S parameter of 1.5 for the full vertical profile, which corresponds to an optical depth of 10 at the surface. The cloud scattering model employed the Mie implementation by Bohren and Huffman, with 20 angles, 200 size bins, and a complex extinction coefficient of 1; for more details see Section 3 of Chapter 5 of the PSG handbook\footnote{https://psg.gsfc.nasa.gov/help.php\#handbook}. Partially cloudy spectra can then be produced by linearly combining the clear and cloudy spectra according to a given cloudiness fraction. While the cloudiness fraction is not a parameter which was optimized during grid construction, it is used as a variable in the retrievals mentioned in Section~\ref{bp}. In all other retrievals, the cloudiness fraction is set to a constant value of 0.5 (meaning that all spectra involved in these retrievals are partially cloudy) unless otherwise stated.

The planetary radius ($R_p$) of each spectrum in the grid was set to 1 $R_\Earth$. Rather than generating spectra with different planetary radii, we took advantage of the proportional relationship between reflected flux and planetary radius for a 1D atmosphere and surface model. A given geometric albedo spectrum from the grid can be converted to a planet-star contrast spectrum with any given planetary radius following Equation~\ref{eq:flux} below. Equation~\ref{eq:flux} shows the planet-star flux ratio $\frac{F_p(\lambda)}{F_s(\lambda)}$ as a function of the geometric albedo spectrum $A_{g}(\lambda)$, planetary radius $R_p$, and planet-star separation $r$ (1 AU throughout this work):

\begin{equation} \label{eq:flux}
\frac{F_p(\lambda)}{F_s(\lambda)} = A_{g}(\lambda) \Phi(\alpha) \left (\frac{R_p}{r} \right)^2, 
\end{equation}

\begin{figure}[H]
\plotone{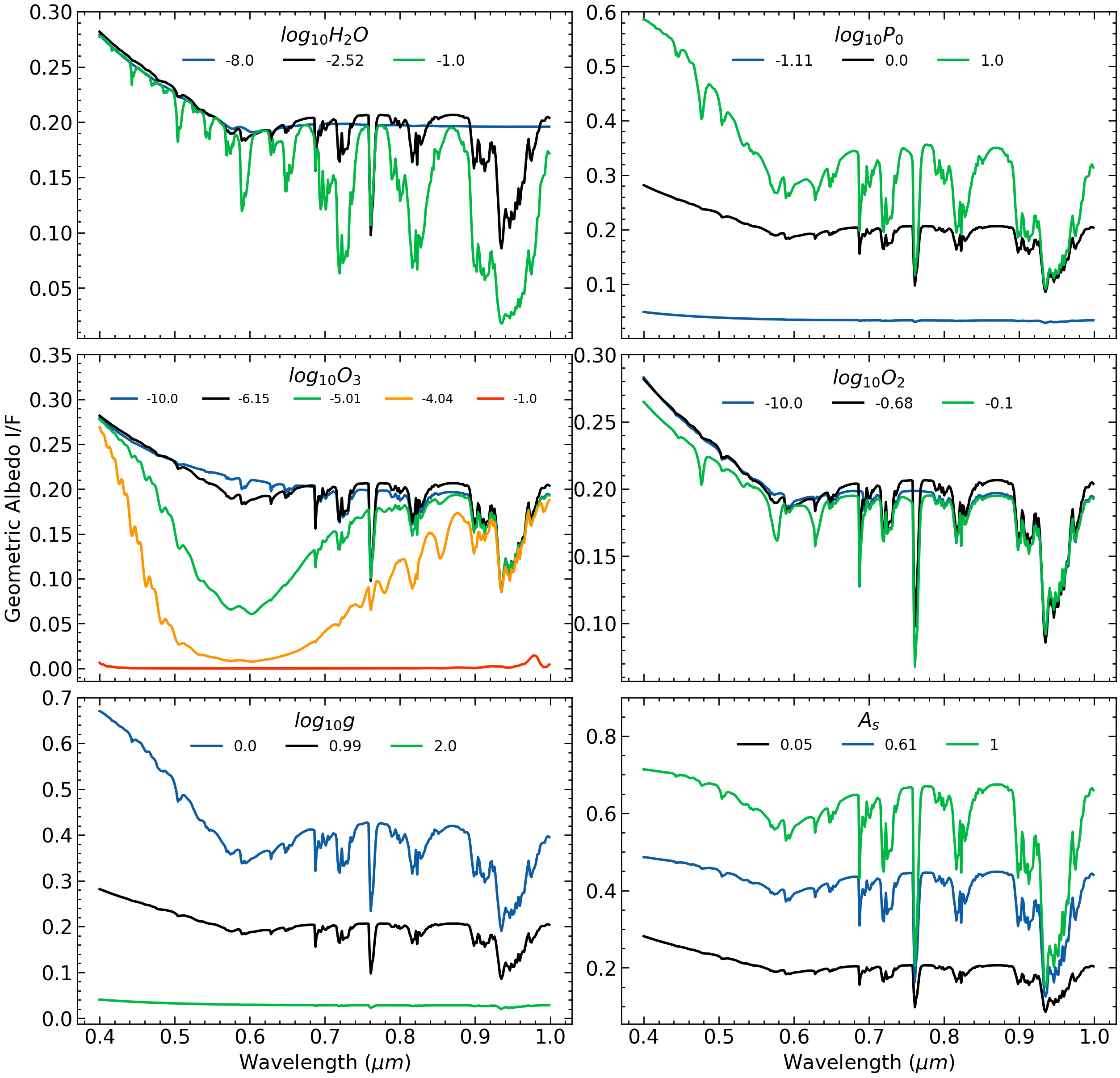}
\caption{Grid spectra at extreme parameter values compared to the fiducial baseline spectrum ($log_{10}H_2O=-2.52, log_{10}P_0=0.0, log_{10}O_3=-6.15, log_{10}O_2=-0.68, log_{10}g=0.99, A_s=0.05$).} In each subplot, only the parameter noted in the upper-center legend is changed. The lower $log_{10}P_0$ spectrum was set to $-1.11$ rather than the grid's minimum extreme of $-3$ because spectra with $log_{10}P_0 \lessapprox -1.11$ appear as flat lines with geometric albedo $\approx 0$.
\label{fig:spectra}
\end{figure}

 
 The spectra are parameterized by their unique $H_2O$, $O_3$, and $O_2$ abundances as well as their surface pressure ($P_0$), gravity ($g$), and surface albedo ($A_s$). These parameter ranges match the prior ranges explored by  \citet[][F18]{2018AJ....155..200F}; the only difference in these parameter ranges is that we cap $P_0$ at 10 bars because greater surface pressures would lead to a complex contribution from clouds, and we deemed this regime to be beyond the scope of the current analysis. The full ranges of these parameters are provided in Table~\ref{tab:parameterization}. Fiducial values were chosen to be the same Earth-like values used in F18: $H_2O=3\times10^{-3}, P_0=1, O_3=7\times10^{-7}, O_2=0.21, g=9.8, A_s=0.05$. Figure ~\ref{fig:spectra} shows the fiducial spectrum in comparison to spectra with extreme high/low values of each parameter. \par 

\begin{deluxetable}{cccc} \label{param_table}
\tablecaption{Grid Parameterization}
\label{tab:parameterization}
\tablewidth{0pt}
\tablehead{
\colhead{Parameter Symbol} & \colhead{Description} & \colhead{Minimum} & \colhead{Maximum}
}
\startdata
$H_2O$ & Water Abundance & $10^{-8}$ & $10^{-1}$ \\
$O_3$ & Ozone Abundance & $10^{-10}$ & $10^{-1}$ \\
$O_2$ & Oxygen Abundance & $10^{-8}$ & $0.8$ \\
$P_0$ & Surface Pressure (bars) & $10^{-3}$ & $10$ \\
$g$ & Surface Gravity ($m/s^2$) & $1$ & $100$ \\
$A_s$ & Surface Albedo & $10^{-2}$ & $1$ \\
\enddata
\end{deluxetable}


\subsection{Grid Construction} \label{construction}
We developed an algorithm to find the optimal spacing and quantity of grid points in order to optimize the trade-off between interpolation accuracy and the computational time needed to produce the grid. For each parameter, we generated a test grid with 75 points for that parameter and 3 points (low, medium and high) for the other parameters. Multiple values of the other parameters are taken into account to capture the effects of all parameters on each spectrum, in order to give a more complete view of how interpolations will be performed in a real retrieval scenario. This leads to a total of $75*3^5=18,225$ grid points for each parameter analysis. The grid point selections for each parameter are handled individually. The algorithm is initialized with a two-point grid composed of one point at each of the upper and lower extrema. This simple grid is then used to interpolate spectra across the full space and the interpolation error (w.r.t. the true spectra in the test grid) is calculated at each point. A new grid point is then added at the location of maximum error from among the 75 test grid points of the parameter of interest. This process is repeated to iteratively add points to the grid at the location of the maximum interpolation error until this maximum error falls below a threshold value.


We utilized an error metric designed to capture the difference between the true and interpolated spectra in proportion to the true spectrum, calculated as:


\begin{equation} \label{eq: error1}
error = \frac{\tilde{d^*}}{max(\bar{s_f},\bar{s_t})}
\end{equation}

where $s_f$ is the fiducial spectrum and $s_t$ is the true spectrum. $\tilde{d^*}$ is the median of the top 10\% of spectral points with the greatest squared difference $d$, which is defined by:

\begin{equation} \label{eq: error2}
d = (s_i - s_t)^2
\end{equation}

\noindent where $s_i$ is the interpolated spectrum and $s_t$ is the true spectrum.

We take the squared difference to treat deviations of the interpolated spectrum above and below the true spectrum the same and to emphasize larger differences over smaller ones. We only consider the 10\% of spectral points with the highest error to further focus on the regions where interpolation is least accurate. The median is then taken over these differences to summarize them as one number, robust to outliers. The fiducial spectrum, which is one with Earth-like parameters as described in \cite{2018AJ....155..200F} ($H_2O=3\times10^{-3}, P_0=1, O_3=7\times10^{-7}, O_2=0.21, g=9.8, A_s=0.05$), was used in the denominator of this equation because spectra which have high $O_3$ abundances have continua near zero, causing the error value to increase substantially (see the $O_3$ panel of Figure~\ref{fig:spectra}). Once a series of grid points that resulted in a maximum interpolation error metric of less than 10\% was found for each parameter, the full grid could be constructed. 

\begin{figure}[H]
\plotone{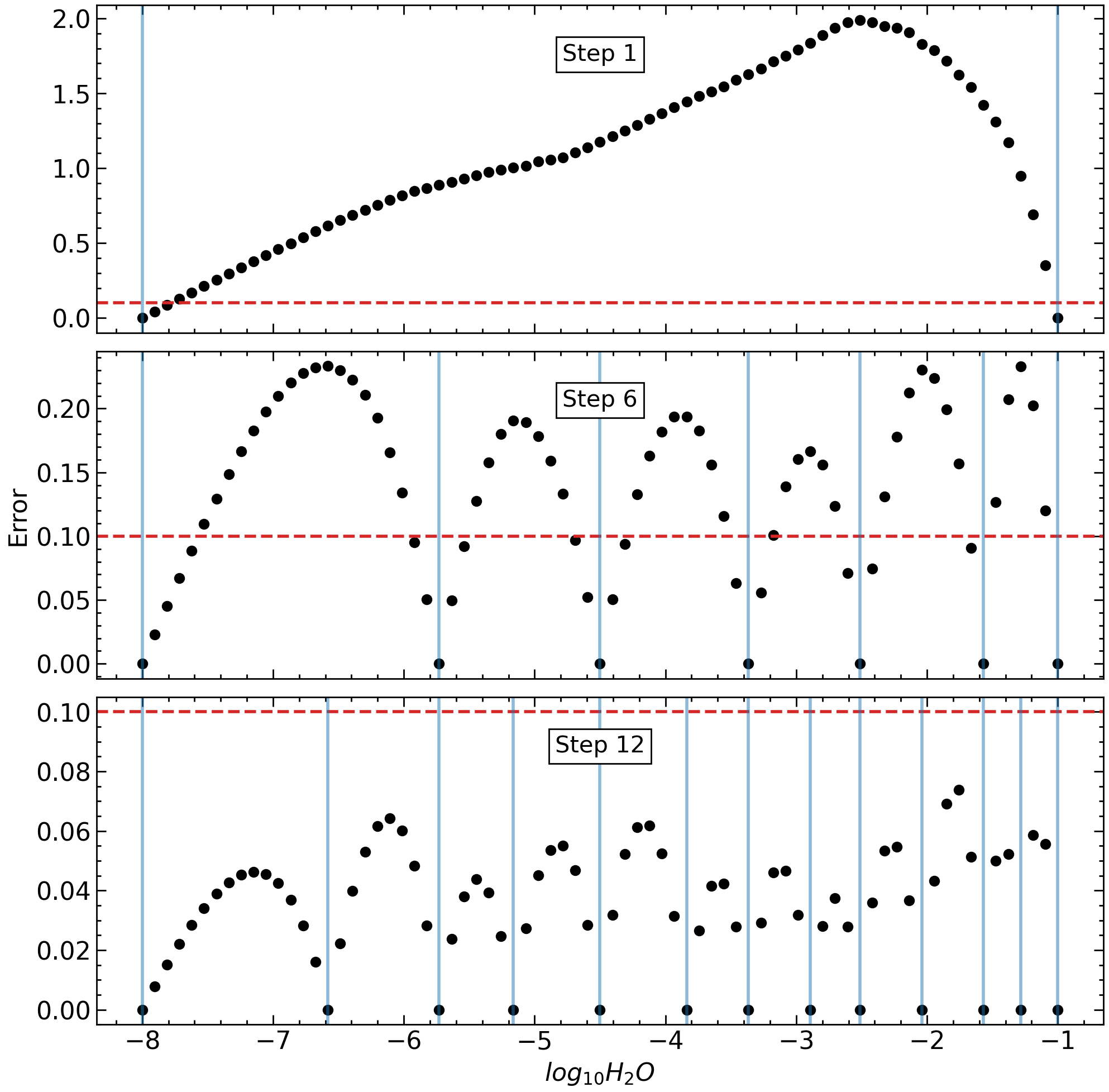}
\caption{Three steps in the grid construction process for $H_2O$. Each step adds one grid point, with step 1 beginning with 2 grid points. The first panel shows the initial state (step 1), the second an intermediate (step 6), and the third panels shows the final grid point configuration (step 12). Each grid point is represented by a blue vertical line. Each black point is the maximum error for the particular $H_2O$ value across all other off-axis parameter values. The horizontal red dashed line is the 10\% error threshold. }
\label{fig:grid_construction}
\end{figure}

Figure~\ref{fig:grid_construction} illustrates the grid construction algorithm with this error metric. As expected, the error is lowest closest to the grid point and on the grid points the error is zero. The error is not necessarily the highest directly in the middle of two grid points (see top panel), so a naive strategy of only placing new grid points in between existing ones is insufficient. 

Our final full grid is composed has 13 points for $H_2O$, 16 for $P_0$, 24 for $O_3$, 9 for $O_2$, 8 for $g$, and 4 for $A_s$ totalling 1.4 million spectra. In order to efficiently generate this large number of spectra, we used a scalable cloud-based interface to PSG housed on GSFC's local cloud computing cluster, which we call GridRunner. GridRunner is able to accelerate calls to PSG's API using in-RAM filesystms and automated configuration of PSG packages. We used GridRunner with about 50 virtual machines, each of which had four workers using 2.6 Ghz Intel Sandy Bridge hardware. This architecture enabled us to generate 1.4 million spectra in about two weeks. The final structure of the grid is summarized in Table~\ref{tab:structure}.\par

\begin{deluxetable}{cc} \label{tab:structure}
\tablecaption{Grid Structure}
\tablewidth{0pt}
\tablehead{
\colhead{Parameter Symbol} & \colhead{Number of Points}
}
\startdata
$H_2O$ & 13 \\
$O_3$ & 24 \\
$O_2$ & 9 \\
$P_0$ & 16 \\
$g$ & 8 \\
$A_s$ & 4 \\
\enddata
\tablecomments{Grid structure (spacing and number of points).}
\end{deluxetable}

\section{Implementation and Validation of Grid-Based Retrievals} \label{retrievals}
\subsection{Choosing a Bayesian Inference Algorithm} \label{sampler}
A key consideration while building a grid-based retrieval framework was the choice of a Bayesian posterior sampling algorithm. This can greatly affect the performance of retrievals so we investigated two popular implementations of two common classes of algorithms, {\tt emcee} \citep{2013PASP..125..306F} for Markov Chain Monte Carlo (MCMC) and Multinest \citep{2008MNRAS.384..449F} for Nested Sampling (NS). While these algorithms both approximate the parameter posterior distributions, the means by which they explore the prior space to infer the distribution differs. An MCMC chain starts from an (often random) initial point  in the prior space and ``walks" to another with a higher likelihood following a stochastic acceptance rule. This continues until a convergence criterion is reached \citep{10.1214/ss/1177011136, 10.1214/20-BA1221} and the paths taken by the walkers are used to construct the posterior distributions. Similarly, NS is initialized by randomly placing a number of \textit{live points} throughout the prior space. A likelihood value is calculated for each of these and the live point with the lowest log-likelihood is discarded. A new point, unrelated to the previous, with higher log-likelihood is then sampled from the restricted prior volume. This process continues until the remaining prior volume is negligibly small, with each discarded live point composing the posterior distribution according to a given weight. 

We examined MCMC and NS through two popular implementations of these methods: {\tt emcee} and Multinest. These packages have been used throughout the literature in the context of exoplanet atmospheric retrievals \citep[e.g.,][]{2016AJ....152..217L, 2018AJ....155..200F, 2018AAS...23114843M, 2022BAAS...54e.081K}. Some works have pointed out weaknesses in these two particular sampling algorithms \citep[in particular Multinest, e.g.,][]{Buchner2016StatisticalTestNestedSampling, 2022A&A...662A.108A, Himes2022PhD}. This work is not intended to be an endorsement of these particular implementations over others; these two were chosen for ease of use and their extensive use by previous studies. Particular results presented here may change slightly when using other sampling algorithms, but the overall methodology presented in this paper can be used independent of the choice of Bayesian framework, and in future implementations we will conduct a more extensive analysis to determine if other sampling algorithms would produce improvements in our results.

We compared {\tt emcee} and Multinest by performing a series of retrievals using both algorithms (the same as those discussed in Section~\ref{psgnest} -- F18's Figures 7-9). In our testing, the final results between the two algorithms were comparable, but MultiNest converged much faster and in far fewer iterations than {\tt emcee}. Specifically, the 4-parameter retrieval discussed in Figure~\ref{fig:PSGnest_fig8} took about 6 seconds to run using our implementation of Multinest (discussed below) and 476 seconds to run using {\tt emcee} (while ensuring convergence according to the the method of \citet{2021arXiv210910843V}). For this reason, we chose to continue this work with Multinest and a NS framework as our Bayesian sampling algorithm.

The choice of whether to interpolate a parameter in linear space or log space can also significantly change the accuracy of the interpolation. To determine the optimal choice for each parameter, we interpolated each parameter individually at multiple values with both spacings and chose the configuration that led to the lowest interpolation error. This led to the choice to interpolate $H_2O$, $P_0$, $O_3$, $O_2$, and $g$ in log$_{10}$ space and $A_s$ in linear space.

\subsection{Development and Benchmarking of PSGnest} \label{psgnest}

We used a novel application of the Planetary Spectrum Generator \citep{2018JQSRT.217...86V, 2022fpsg.book.....V} called PSGnest\footnote{https://psg.gsfc.nasa.gov/apps/psgnest.php} to perform the data analysis shown in this paper. PSGnest is a retrieval tool based on Multinest which is specifically conceived for exoplanetary observations, yet it can be adapted to any data with the proper setup. PSGnest takes advantage of memory mapping methods coded in C to greatly accelerate computations. PSGnest outputs all the relevant quantities for nested sampling, including the log evidence log(Z), the highest-likelihood output parameters, their average value resulting from the (possibly multimodal) posterior distribution and their uncertainties, which are estimated from the posterior distribution as well \citep{2022fpsg.book.....V}. Unless otherwise stated, all of the retrievals presented in this work used the PSGNest default values for certain MultiNest hyperparameters. These include 400 live points (examined further in Figure~\ref{fig:retrieval_time}) and a stopping/convergence factor (dlogz) of 0.1. The sampling efficiency factor was set to 1.0, favoring posterior accuracy over evidence accuracy. Constant efficiency mode, which is known to underestimate errorbounds, was not used.

\begin{figure}[H]
\includegraphics[scale=0.8]{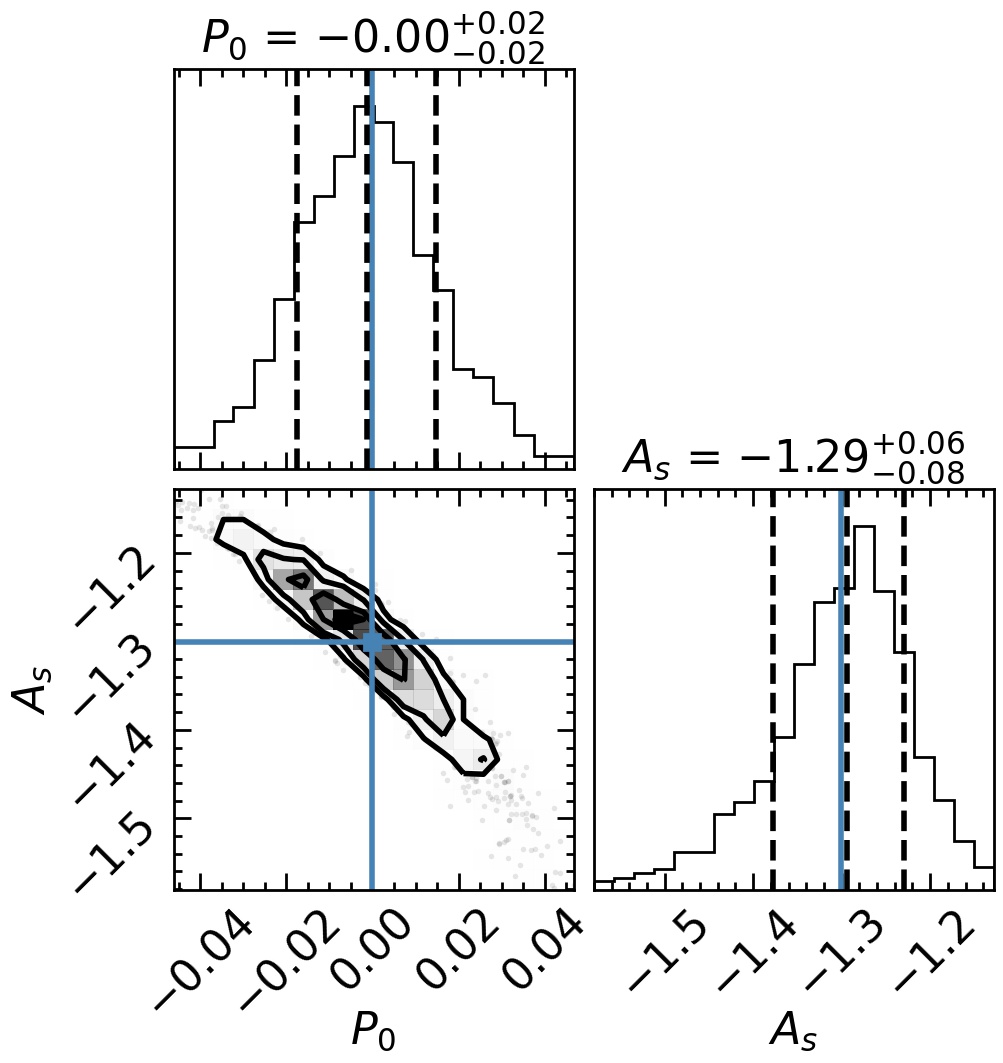}
\caption{Reproduction of F18's Figure 7 using PSGnest.  \label{fig:PSGnest_fig7}}
\end{figure}

\begin{figure}[H]
\plotone{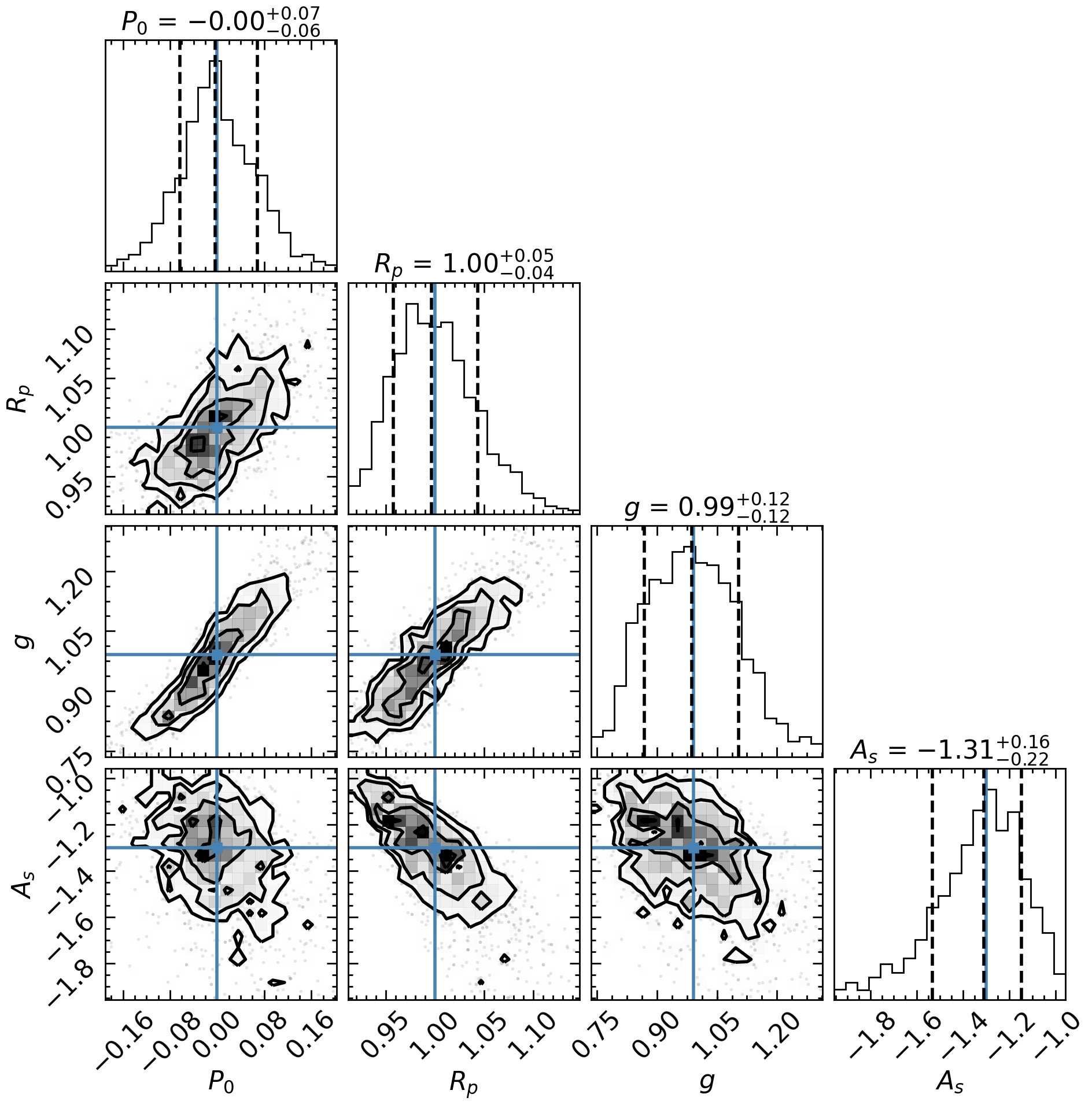}
\caption{Reproduction of F18's Figure 8 using PSGnest.  \label{fig:PSGnest_fig8}}
\end{figure}

\begin{figure}[H]
\plotone{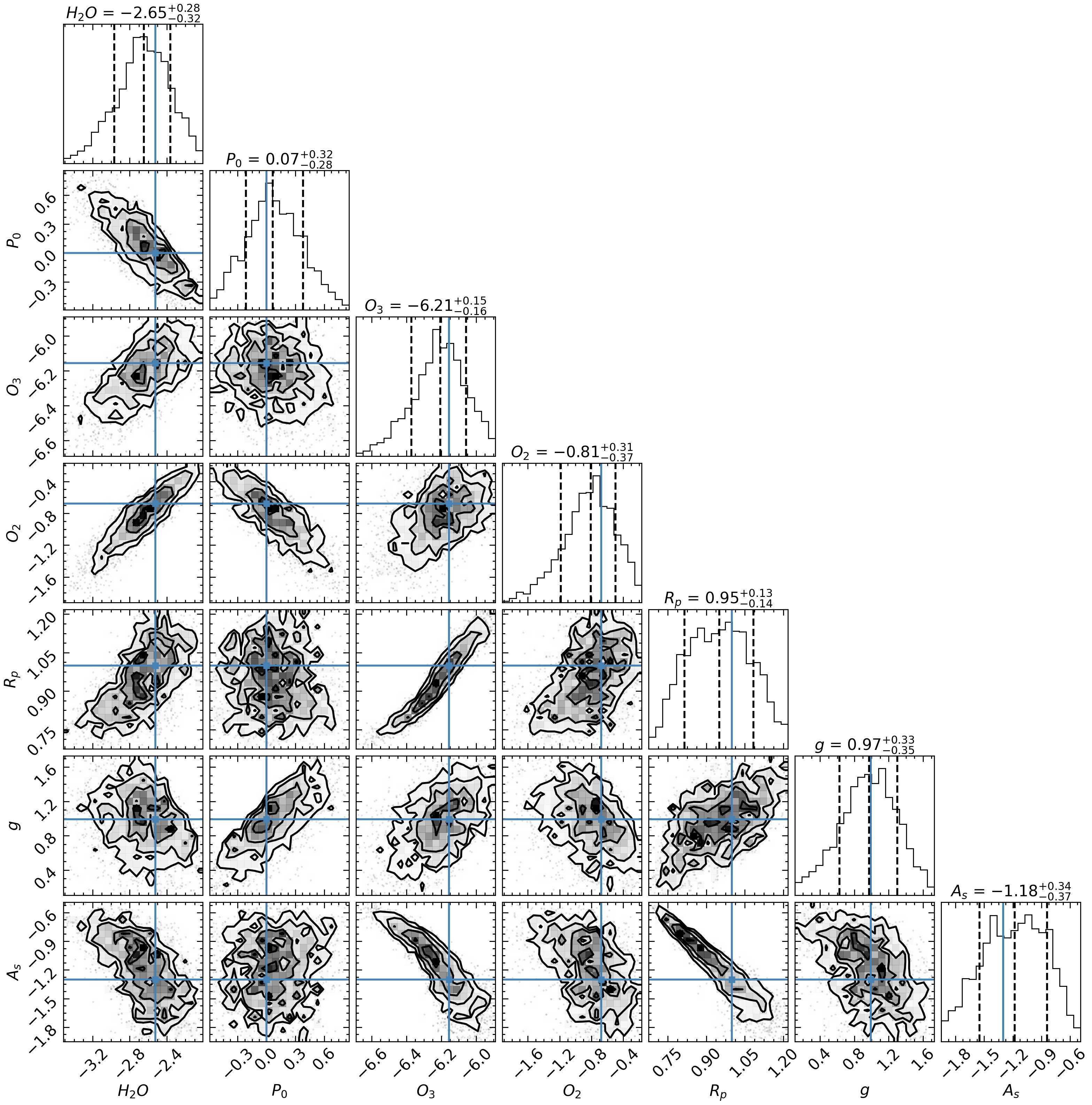}
\caption{Reproduction of F18's Figure 9 using PSGnest.  \label{fig:PSGnest_fig9}}
\end{figure}

\begin{figure}[H]
\plotone{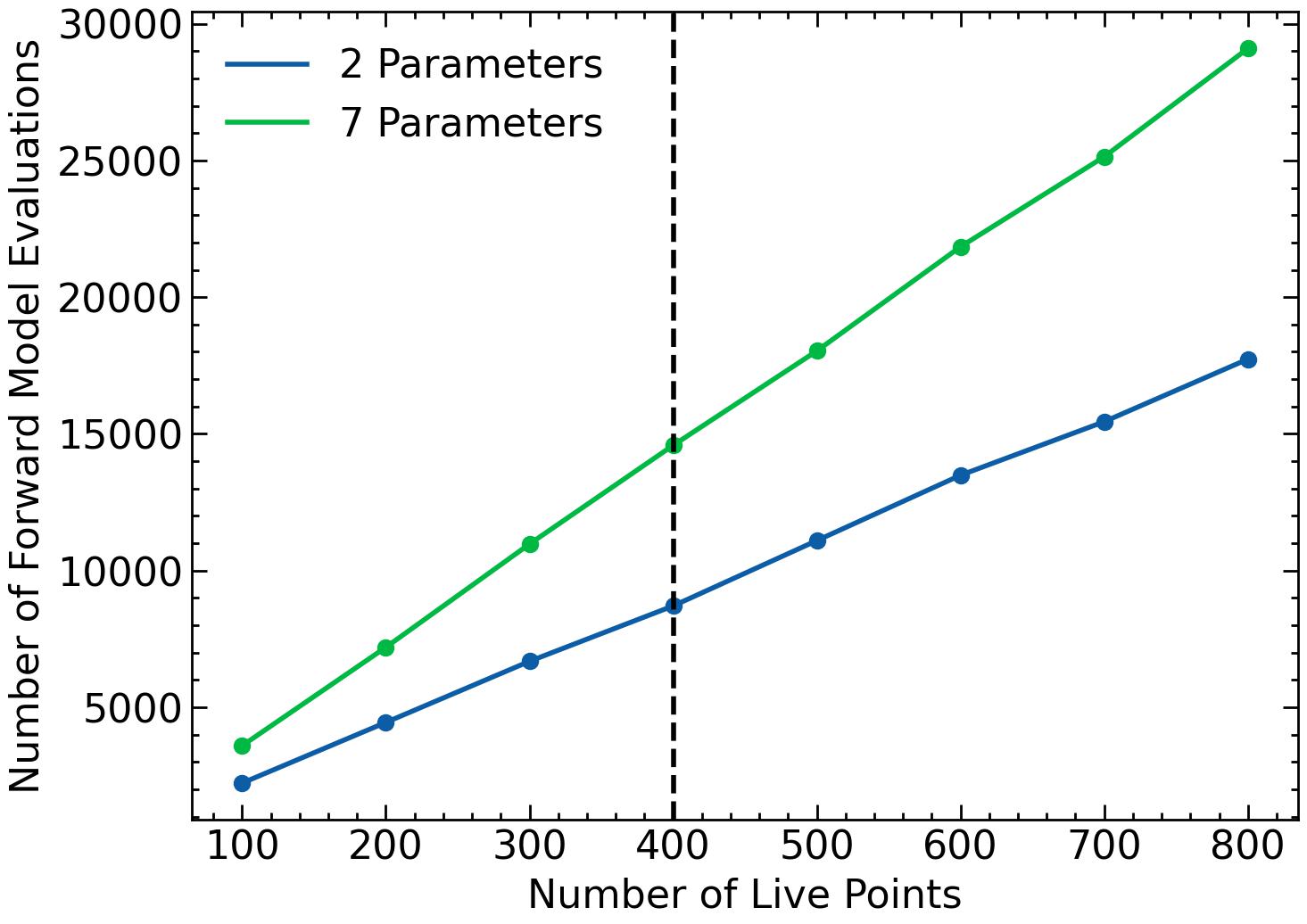}
\caption{Number of forward model evaluations of PSGnest for 2- (F18's Figure 7) and 7- (F18's Figure 9) parameter retrievals by number of live points}. The number of live points is directly correlated with the completeness of the posterior's sampling but too many can increase the runtime of the retrieval (proxied here by the number of forward model evaluations) without significantly changing retrieval results. We found that 400 live points (marked with a vertical dashed line) is enough to sufficiently sample the posteriors and a number of live points beyond this does not noticeably change the results.
\label{fig:retrieval_time}
\end{figure}

Once the grid was constructed, we proceeded to use it to benchmark PSGnest. This was primarily done by performing retrievals configured in a manner similar to those described in F18. There are some differences between the radiative transfer scheme used here versus that of F18, mainly that we use isotropic clouds while F18 defines a distinct cloud layer. Figures 7, 8, and 9 of F18 show their retrieval results for retrievals of 2 ($P_0$, $A_s$), 4 ($P_0$, $R_p$, $g$, $A_s$), and 7 ($H_2O$, $P_0$, $O_3$, $O_2$, $R_p$, $g$, $A_s$) parameters, all parameters included in our grid (except for $R_p$, see Section~\ref{construction}). We perform retrievals to reproduce these figures using spectra interpolated from our grid as the forward model to the PSGnest retrieval interface. The goal of this study is to benchmark the performance of PSGnest through comparison to ensure that it is a correctly-implemented interface to MultiNest. To this end, we set the data spectra of the retrievals to be a spectrum interpolated from the grid. Thus, interpolation error will not confound the sampler's ability to compare model spectra to the data spectrum and this study will only investigate the validity of PSGnest, independent from the interpolation error from the grid. The spectra were converted from geometric albedo to star/planet contrast following Equation~\ref{eq:flux}, which enables us to include the planetary radius, $R_p$, in the retrieval (see Section~\ref{construction}). In addition to this, the spectra of the grid were downgraded from R=500 to R=140 by averaging together the hi-res spectral values closest to each low-res wavelength point to match F18, and S/N=20 uncertainty was applied during the retrieval (without random scatter, see Section~\ref{scatter}).  We used partly cloudy spectra (computed by adding 50\% of a clear spectrum to 50\% of a cloudy spectrum). The results of this reproduction are shown in the Figures~\ref{fig:PSGnest_fig7}, ~\ref{fig:PSGnest_fig8}, and ~\ref{fig:PSGnest_fig9}.

Overall, we find that the results produced by PSGnest compare favorably to those presented in F18. The retrieved values and uncertainties are similar to those of F18. Differences are likely due to the grid-based nature of PSGnest (although interpolation error does not factor this) and the difference in sampling techniques. PSGnest is built on MultiNest \citep{2008MNRAS.384..449F} whereas F18 used {\tt emcee} \citep{2013PASP..125..306F}. We found that PSGnest/Multinest performs much faster than {\tt emcee}, with PSGnest taking on the order of minutes  while {\tt emcee} took hours if care is taken to ensure the sampler converges to a solution. Figure~\ref{fig:retrieval_time} summarizes the number of forward model evaluations used by PSGnest for different numbers of live points. These retrievals, as well as all others discussed in this work, were run on a 2020 MacBook Pro with a 2.0GHz quad‑core 10th‑generation Intel Core i5 processor \par

One area where noticeable differences are present in our retrievals compared to those of F18 is in the correlations between parameters, particularly in the reproduction of their Figure 9 (our Figure~\ref{fig:PSGnest_fig9}). We observe a weak anticorrelation between $H_2O$ and $P_0$ and also $O_2$ and $P_0$ whereas the respective figure in F18 shows positive correlations. We are confident in the validity of our findings due to physical expectations. High $P_0$ leads to deeper spectral features for the same gas abundance, so matching a spectrum with higher $P_0$ requires a lower abundance. Therefore, some degree of anticorrelation should be expected. 

\subsection{Grid Evaluation} \label{eval}
Before applying the grid towards a scientific study, we sought to characterize the error induced by interpolating spectra between grid points and its effect on retrievals across the parameter space. To this end, we generated a ``test grid" composed of actual PSG-derived spectra with grid parameter values chosen to be intermediate between the grid point positions of the original grid of spectra, which we refer to as the ``main" grid. The test grid spectra were generated in a manner that is otherwise identical to the main grid's spectra (see Section~\ref{construction}). This test grid allows us to explore the interpolation error at the locations where it should theoretically be the highest, thus yielding a worst-case estimation of the error. 

\subsubsection{Interpolation Error} \label{interp err sec} 
\begin{figure}[H]
\plotone{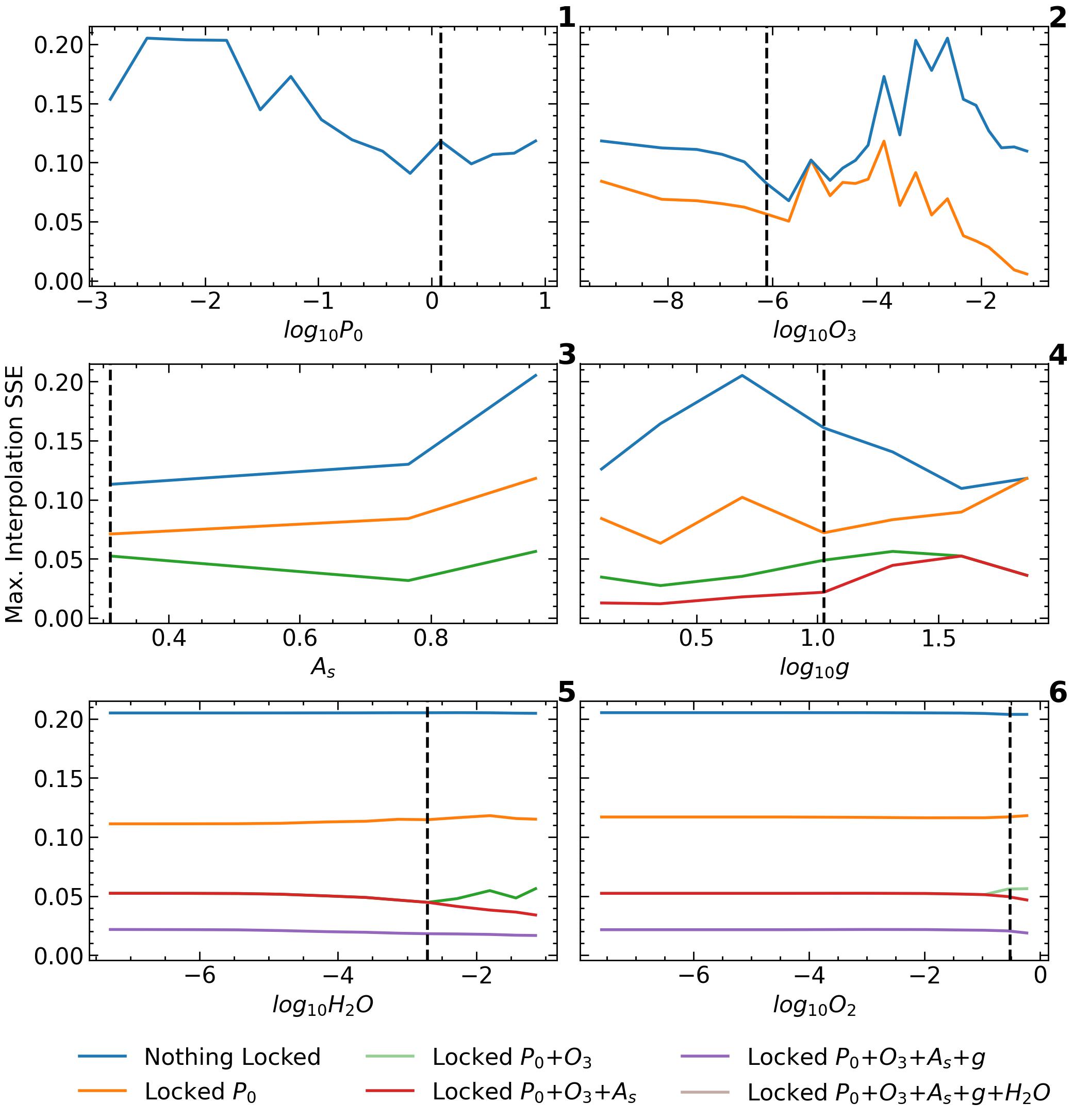}
\caption{Interpolation error across the parameter space. The blue curves represent the maximum interpolation error at each point in the parameter space over all off-axis parameters. The non-blue curves are created by sequentially filtering out all values other than an Earth-like value for each respective parameter (other parameters are not filtered unless noted). The locked parameters are set to Earth-like values (also listed in Table~\ref{tab:earth-like_vals}) of 0.002 for $H_2O$, 1.205 for $P_0$, 7.796$*10^{-7}$ for $O_3$, 0.301 for $O_2$, 10.642 for $g$, and $0.311$ for $A_s$, and are represented by vertical black dashed lines. These are points on the test grid which are closest to realistic Earth-like values. The parameters are locked in descending order of impact on the maximum interpolation error SSE. Panel 1 is the maximum interpolation SSE with no parameters locked; Panel 2 adds the effect of locking $P_0$; Panel 3 adds the effect of locking $O_3$; Panels 4, 5, and 6 add the effects of locking $A_s$, $g$, and $H_2O$ respectively (the maximum interpolation SEE with $P_0$, $O_3$, $A_s$, $g$, and $H_2O$ locked is entirely in line with the curve with $P_0$, $O_3$, $A_s$ and $g$ locked).
\label{fig:interp err}}
\end{figure}

First, we calculated the interpolation error for each of the 700,000 test grid spectra using two metrics: the sum of squared errors (SSE) between the true and interpolated spectrum, and the coefficient of determination ($R{^2}$), which measures the linear correlation between each interpolated and true spectra. As a first check for the performance of the grid, we examined the mean $R^2$ across the entire grid space and found this value to be 0.998, indicating very accurate interpolations on average. To look closer into where the interpolation accuracy is poor, we took the maximum of the interpolation SSE's for all spectra of each parameter value of the six grid parameters. This yielded the highest interpolation error across the parameter space for every distinct value of every parameter and is shown in Figure~\ref{fig:interp err}.

\begin{deluxetable}{cc} \label{tab:earth-like_vals}
\tablecaption{Earth-like off-axis values}
\label{tab:earthlike}
\tablewidth{0pt}
\tablehead{
\colhead{Parameter Symbol} & \colhead{Value}
}
\startdata
$H_2O$ & 0.002 \\
$O_3$ & 7.796$*10^{-7}$ \\
$O_2$ & 0.301 \\
$P_0$ & 1.205 \\
$g$ & 10.642 \\
$A_s$ & 0.311 \\
\enddata
\tablecomments{Earth-like values used for off-axis parameters throughout this study. These were chosen as the points on the test grid which are closest to the Earth-like values used in F18.}
\end{deluxetable}

In Figure~\ref{fig:interp err} we plot the highest interpolation error values for each parameter. The blue curves in Figure~\ref{fig:interp err} represent the maximum interpolation error for each parameter across all values of the off-axis parameters. We observe interpolation errors which are high relative to the rest of the respective parameter space in regions of low $P_0$ ($log_{10}P_0<-1.5$), high $O_3$ ($-5<log_{10}O_3<-1.5$), high $A_s$ ($A_s \sim 0.9$), and moderate $g$ ($log_{10}g \sim 0.6$). The overall interpolation error does not vary significantly as a function of $H_2O$ and $O_2$. Retrievals of observed exoplanetary spectra with planetary parameters in these regions of high interpolation error could suffer from poorer accuracy than retrievals of spectra with atmospheric parameters in a region with low interpolation error, but this is dependent on the relative impact of interpolation error versus the contribution from noise in the data and the degeneracies due to the combination of parameters retrieved.

A key point is that the accuracy of interpolating between values of a single parameter is highly dependent on the values of the off-axis parameters. Each of the non-blue curves in Figure~\ref{fig:interp err} shows the interpolation error when a given off-axis parameter is locked to a single, Earth-like value on the test grid (see Table~\ref{tab:earthlike}). Locking an off-axis parameter to a single value allows us to estimate the effect that the parameter has on the interpolation error overall by comparing these narrower interpolations to those performed across every off-axis parameter value. We chose to focus on the Earth-like parameter space for this examination because we consider this to be an important planetary regime for the purpose of this study, and it is clear from this same figure that the grid interpolation works relatively well in this regime. The non-blue curves show that the interpolation error in the Earth-like regime produces a significantly lower interpolation error than the worst portions of parameter space. For instance, we observe in panel 2 (upper right) of Figure~\ref{fig:interp err} that locking $P_0$ at the Earth-like value reduces the maximum interpolation SSE of $O_3$ from 0.2 to 0.1. From this, we can conclude that $P_0$ has a significant impact on the accuracy of interpolation between $O_3$ values. This trend continues for each parameter - we see that locking $P_0$ has the greatest impact on the maximum interpolation errors of each other parameter, followed by $O_3$, then $A_s$. Locking $g$, $H_2O$, or $O_2$ has a much smaller effect on the interpolation error than locking any other parameter. Therefore, $P_0$, $O_3$, and $A_s$ are the primary drivers of the interpolation error in this grid. 

\subsubsection{Retrieval Performance in the Earth-like Regime} \label{retr err sec}

Once the interpolation error across the parameter space was characterized, our next step was to evaluate the grid on its performance when used in retrievals. We ran 6-dimensional retrievals for each distinct value of each parameter in the test grid. For each of these, the parameter of interest was set to be the given value in the test grid and the off-axis parameters were set to be Earth-like values (see Table~\ref{tab:earthlike}). For each of these parameter sets, a ``true data" spectrum was created with an S/N of 20 from the original PSG radiative transfer calculation. The retrievals were then repeated but instead of using a ``true" spectrum for the data, the data spectrum was created by interpolating from the main grid. By comparing the retrieval accuracy between a ``true" data spectrum and an interpolated data spectrum, we can determine the offset in the retrieved values created by the interpolation error, since the retrieval sampling algorithm should be able to match the interpolated data spectrum perfectly but should retrieve a somewhat imperfect best-fit value for the true spectrum. No random noise was added to the data spectra in any of these retrievals because this noise would prevent the retrievals which used an interpolated data spectrum from perfectly matching the input.

\begin{figure}[H]
\plotone{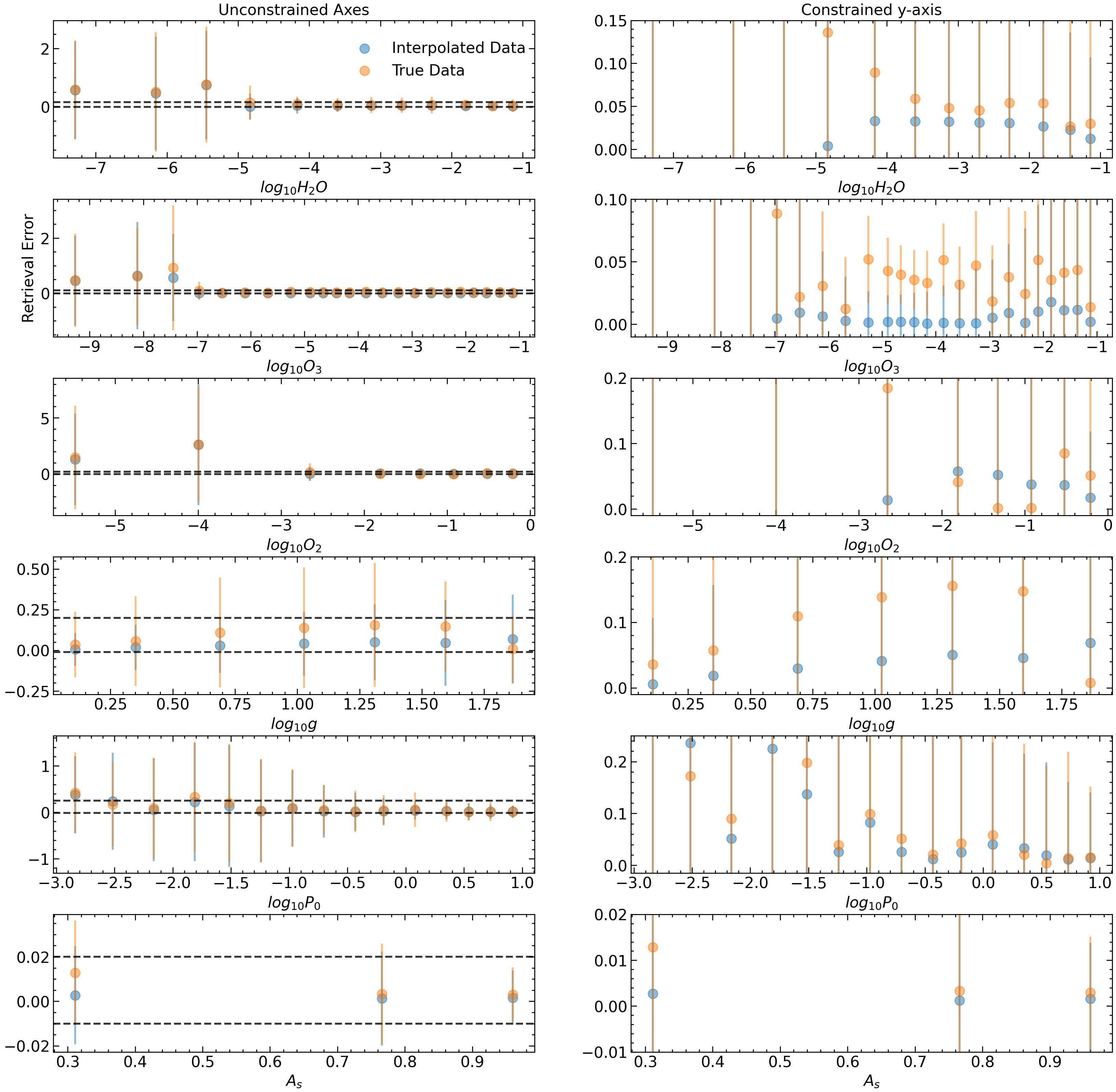}
\caption{6-D Retrieval Evaluations showing the retrieval error across the parameter space. The retrieval error is defined as the absolute difference between the true and retrieved values. Blue points represent retrievals performed using an interpolated data spectrum while orange points represent retrievals performed using true data spectra. Error bars on these points are drawn as the 68\% credible regions. The left column shows the full extent of the y-axes, while the right column zooms into the portion of the y-axes contained within the horizontal dashed black lines. }
\label{fig:retr_eval}
\end{figure}

Figure~\ref{fig:retr_eval} displays the results of these retrievals. In this figure, retrievals performed using the interpolated data spectra are represented in blue and those performed using ``true" data spectra are shown in orange. This figure provides two different views of the retrieval error, calculated as the absolute offset between the retrieved parameter value for a retrieval performed using a ``true" data spectrum and the retrieved value when using an interpolated one. The left column shows the full extent of the offset, while the right column zooms in to regions where the two types of retrievals yielded similar offsets. By comparing the interpolated data results to the ``true" data results and examining how the difference compares to the uncertainty driven by the retrieval analysis, we can examine the effect of the interpolation error on the derived best-fit value while other confounding factors such as degeneracies between parameters and uncertainty due to the data uncertainties are controlled. Any differences between the results of the two series of retrievals are due to the interpolation error. 

As we would expect, the derived retrieval error is either higher for the ``true" data spectrum retrievals, or very similar for both ``true" and interpolated data retrievals. We also find that the derived retrieval error for both the retrievals performed using a true data spectrum and those resulting from the use of an interpolated spectrum are almost always within $1\sigma$ of the input value. The only outliers are the ``true" data spectrum retrievals for several values for O3, which are $\sim1.2\sigma$ from the input value; these points represent the highest-interpolation error points from panel 2 in Figure~\ref{fig:interp err}, and we further discuss the specific reasons for the excess error in these regions of parameter space below. Overall, we conclude the the interpolation error is not a significant inhibitor for performing retrievals of planetary spectra in the region of parameter space close to the Earth-like regime using this grid.

\subsubsection{Retrieval Performance for High-Error Regions} \label{region}
\begin{figure}[H]
\plotone{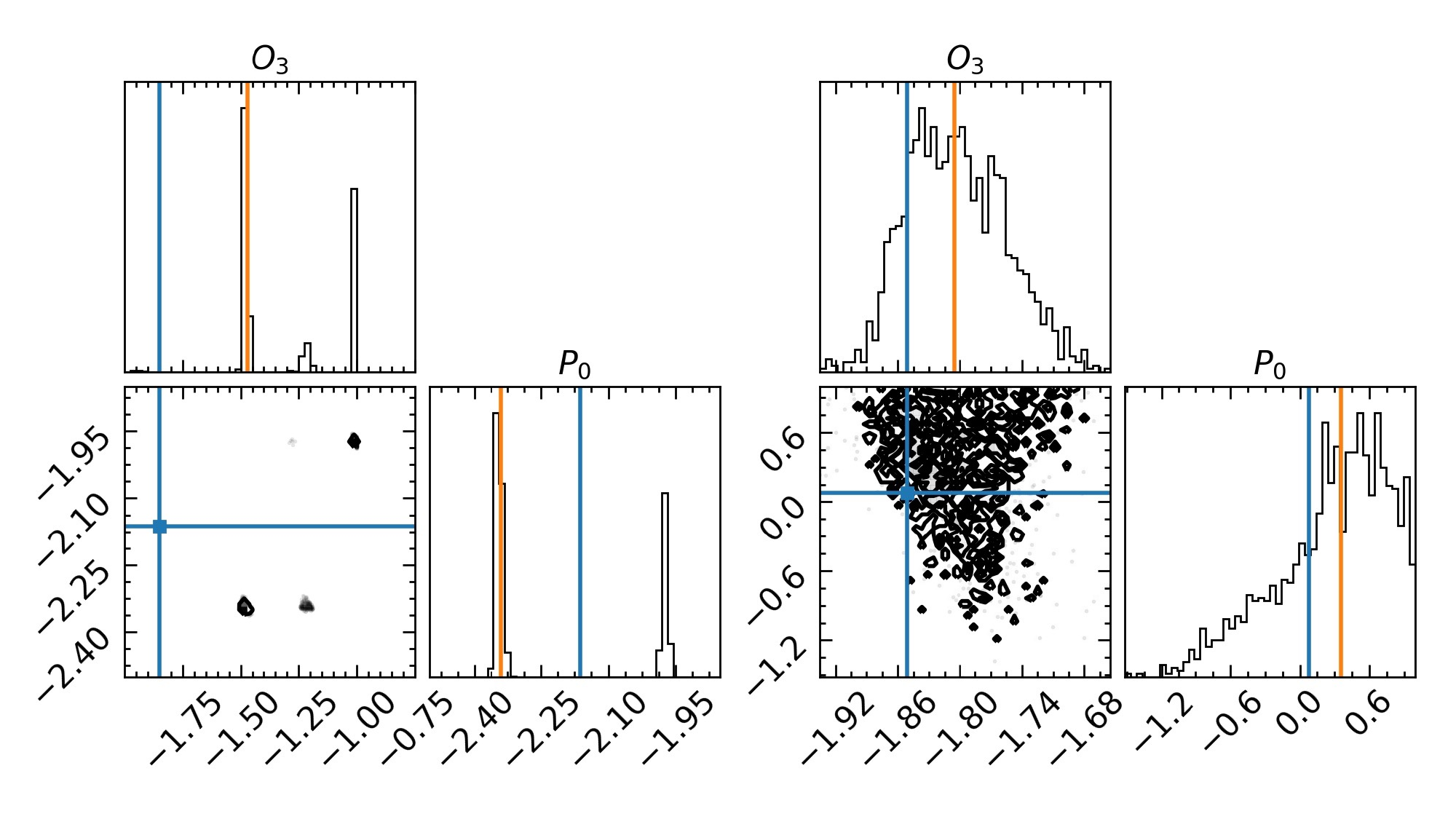}
\caption{2-D posteriors of $O_3$ and $P_0$ resulting from two separate 6-D retrievals of data spectra both with $O_3=0.014$. The corner plot on the left shows a retrieval with a low true surface pressure ($P_0=0.0068$) while the panel on the right shows a retrieval with an Earth-like value for the true surface pressure ($P_0=1.2$). The true and retrieved (median) values are shown as blue and orange vertical lines respectively. The retrieval algorithm fails to recover the correct $O_3$ and $P_0$ and converges to several other solutions when $O_3$ is high and $P_0$ is low but is more accurate when both $O_3$ and $P_0$ are high.}
\label{fig:O3_corner}
\end{figure}

While the evaluation metrics discussed in Sections~\ref{interp err sec} and~\ref{retr err sec} show that the interpolation error does not significantly impair retrievals of spectra in the Earth-like regime (recall Table~\ref{tab:earthlike}), this level of performance does not hold for all regions of the grid's parameter space. We observed remarkably poor retrieval performance (i.e. the retrieved values differ significantly from the true values) in regions of high $O_3$ ($log_{10}O_3>-5$), low $P_0$ ($log_{10}P_0<-1.7$), and high $A_s$ ($A_s=0.96$) where the Bayesian inference converges to incorrect values and around multiple distinct modes appearing as spikes. Figure ~\ref{fig:O3_corner} shows an example of these inaccurate retrievals alongside a more typical case.

This 6-D retrieval of a data spectrum with $O_3=0.014$, $P_0=0.0068$, and $A_s = 0.960$ converges to incorrect solutions focused around several $O_3$ and $P_0$ points. These spikes are also present in the $g$ posterior, but this marginalized parameter does not contribute significantly to the interpolation error (Figure~\ref{fig:interp err}). The spikes are not present in the $H_2O$, $O_2$, and $A_s$ marginalized posteriors associated with these retrievals. To confirm that these results are correlated with regions of high $O_3$ and low $P_0$, we produced a similar corner plot for a retrieval using the same high $O_3$ value ($O_3=0.014$) but with Earth-like $P_0$ and other off-axis parameters (Table~\ref{tab:earthlike}), shown in the right panel. The right corner plot does not exhibit the same problems present in the low $P_0$ retrieval of the left corner plot. We also examined retrievals involving other combinations of high/low $O_3$ and $P_0$ and found that these issues were not present in other cases. Therefore, it is the interaction between high $O_3$ and low $P_0$, not one parameter individually, which causes the degeneracy and poor retrieval results.

We then repeated the retrieval whose results are shown in the left panel of Figure~\ref{fig:O3_corner} but using an interpolated data spectrum instead of a true one produced by PSG (the same approach used in Section~\ref{retr err sec}). This allows us to investigate this effect independent of interpolation error. We observed similar spikes in this case as before, but most solutions were centered near the true values. This indicates that the interpolation error contributes to the significant error in the retrieved best-fit values, but is not the cause of the unusual posterior morphology.  To investigate this further, we plot the true and retrieved spectra as well as the spectra associated with the left, middle, and right points in the 2-D marginalized posterior shown in Figure~\ref{fig:O3_corner}. 

\begin{figure}[H]
\plotone{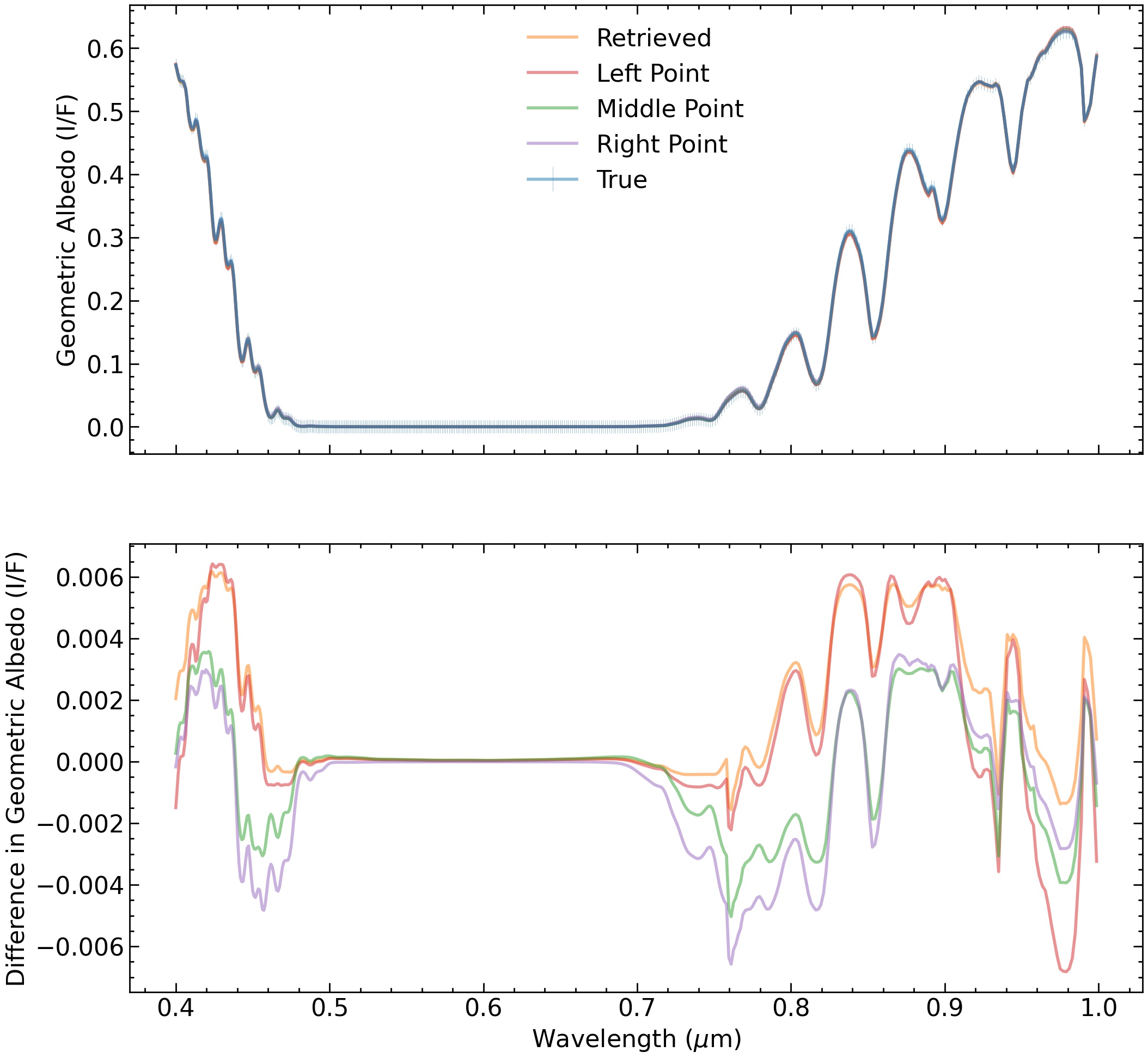}
\caption{Spectra associated with retrievals in the high $O_3$, low $P_0$, high $A_s$ space. "Left", "Middle", and "Right" refer to the three darkest points in the 2-D posterior of the left panel of Figure~\ref{fig:O3_corner}. The true spectrum corresponds to the blue vertical lines in that figure while the retrieved corresponds to the orange. The second panel plots the difference between the true spectrum and each other spectrum. Despite differences in parameter values, all spectra are nearly identical.}
\label{fig:O3_spectra}
\end{figure}

Plotting the true and retrieved spectra alongside other spectra explored by the sampler reveals the degenerate nature of this regime. The lower panel of Figure~\ref{fig:O3_spectra} shows the difference between each spectrum drawn from the posterior and the data spectrum. These differences are relatively minuscule, indicating that each of these spectra are nearly identical despite their different locations in the parameter space. This degenerate behavior may also be inherent to the nature of this particular region of parameter space and the combination of high $O_3$ and low $P_0$ values. These degenerate solutions are particularly acute for atmospheres with high levels of $O_3$. The modeling of $O_3$ in the UV/optical is done in PSG employing cross-sections, which for highly optically thick regimes can be challenging to capture the subtle changes at high opacities on the line cores and wings; this may lead to a relatively high interpolation error compared with the small changes in the spectral shape. Unfortunately, there are no available linelists databases for these bands, so no line-by-line or correlated-k methods can be applied at these wavelengths, which could assist in removing these degeneracies in the modeling.


Examining this issue further, we notice that the positions of these spikes correspond directly to grid points in the case where the data spectrum is not interpolated from the grid (i.e. when interpolation error confounds the sampler's ability to reach the true solution). Since we established in Figure~\ref{fig:O3_spectra} that the spectra in this region are all practically identical, the likelihood values associated with these spectra will also be nearly identical. This will cause the sampler to prefer all degenerate solutions nearly equally. However, because we use grid-based methods where interpolation error causes deviations from the true spectra, interpolated spectra which are closest to the their true versions will be preferred by the sampler because these will be closest to the degenerate solution. This occurs where the interpolation error is the lowest: on the grid points themselves. Therefore, these spikes can be expected when using grid-based methods in a region of high degeneracy.

Adding random scatter to the data spectrum would make this issue less pronounced as the singular true spectrum is altered, but it is unclear if this would fully alleviate the issue; additional work will be needed to further characterize this effect. The results shown here are a worst-scenario for retrievals using this grid.

\section{Application of Grid-Based Retrievals} \label{bp}
With the grid and the PSGnest retrieval framework validated, we proceeded to apply these methods to a scientific case. Extensive studies to understand the potential yields of future direct imaging mission concepts (such as the LUVOIR and HabEx concepts prepared for the Astro2020 Decadal Survey) will be important in designing and optimizing mission architecture and instrumentation. In particular, direct imaging missions incorporating internal coronagraphs to block the light of the central star will be limited in the width of the simultaneous bandpass that can be acquired. Additionally, the use of imaging spectrograph technologies such as an integral field spectrograph (IFS) will limit the spectral resolving power that can be achieved. By comparing the atmospheric constraints achieved with different mission and instrument performance expectations, we can determine the best balance of various aspects of the instrumentation. Similarly, we would like to optimize the total integration time needed to acquire specific constraints on atmospheric parameters, which are driven by the wavelength coverage and S/N of the data being examined.

\subsection{Optimal Bandpass Study Methods} \label{bp methods}
To this end, we applied our grid-based retrieval framework to investigate the potential ``characterization yields", or constraints on different atmospheric parameters, as a function of  different bandpass centers, bandpass widths, S/N, and spectral resolving powers (R). We examined three scenarios for atmospheric bulk density  (i.e. $P_0$ and $g$), in order to inform the parameterizations of the spectra used in this study:  atmospheric surface pressures and surface gravities analogous to those of Mars, Earth, and Neptune. Each of these bulk atmospheric density scenarios has the same gas abundance values as the Earth-like case of F18 (constant VMRs of $H_2O=3*10^{-3}$,  $O_3=7*10^{-7}$, $O_2=0.21$) and surface albedo $A_s$ and planetary radius $R_p$ of a realistic Earth (0.3 and 1 $R_\Earth$ respectively). The planetary scenarios therefore only differ in their surface pressure and gravity, which more closely emulate those of the Solar System planetary analogs. These parameter values are presented in Table~\ref{tab:bulk_parameterization}. Note that the parameterizations for these planetary scenarios were not meant to replicate the respective planets exactly -- only to provide different $P_0$/$g$ archetypes for comparison.

In contrast to the previous retrievals we ran, here the cloudiness fraction is now varying and the planetary radius is fixed at 1 $R_\Earth$. The cloudiness fraction, $C_f$, controls the linear combination of clear ($s_{clear}$) and cloudy ($s_{cloudy}$) spectra: $C_f \times s_{clear} + [1-C_f] \times s_{clear}$. Fixing the planetary radius leads to unrealistically strong constraints on the surface pressure and surface albedo because of the degeneracy between the impact of the planetary radius and the impact of these parameters, but this is necessary since our current retrieval framework does not allow for including physically-based priors on planetary radius with true planetary radii other than 1 $R_\Earth$. Disentangling true $R_p$ values from other factors using only directly imaged reflected-light spectra of planets is essentially impossible, and therefore ancillary constraints are necessary, We leave this type of prior constraint retrieval for future work.  \par

\begin{deluxetable}{cccc} \label{tab:bp_params}
\tablecaption{Bandpass Study Planetary Bulk Parameterizations}
\label{tab:bulk_parameterization}
\tablewidth{0pt}
\tablehead{
\colhead{Parameter Symbol} & \colhead{Mars-like} & \colhead{Earth-like} & \colhead{Neptune-like}
}
\startdata
$P_0$ (bars) & 0.00636  & 1.0 & 10 \\
$g$ ($m/s^2$) & 3.71  & 9.8 & 11.15 \\
\enddata
\tablecomments{Values of $P_0$ and $g$ for the planetary scenarios. All other parameter are the same for each scenario. The $P_0$ upper limit for the Neptune-like case is restricted by the parameter space of the grid (recall Table~\ref{tab:parameterization})}
\end{deluxetable}

We chose a fiducial parameter set with a bandwidth of 10\%, S/N$=$10, and R$=$140 to be a baseline for comparison; this is similar to the values assumed in the LUVOIR concept study for initial characterization measurements of Earth-like planets \citep{2019arXiv191206219T}. From there, we varied one of the bandwidth, S/N, or resolving power to explore the sensitivity of the constraints to these instrumental factors. A bandwidth of 20\%, R of 90, and S/N of 20 were individually adopted for each planetary scenario.  To test the impact of bandpass position, we chose 25 evenly-spaced bandpass center positions within the wavelength range of 0.515--1.0 $\mathrm{\mu}$m, which matches the visible wavelength range of the LUVOIR mission concept \citep{2021AJ....161..150C}. The 20\% bandwidth cases used the same bandpass centers as the 10\% cases but with any bandpasses extending beyond the 0.515-1.0 $\mathrm{\mu}$m range removed. We took portions of the planetary analog spectra around these center positions corresponding to a given fractional bandwidth.  We then ran a retrieval with the given slice of the planetary analog spectrum as the data spectrum. For each of the 25 retrievals performed, we record the median value and upper and lower limits of the 68\% credible region  of the posteriors \citep[following the recommendations of][]{2022PSJ.....3...80H}. We calculate the Bayes factor for each retrieval run in order to confirm that the constraint achieved on the gaseous parameters can be considered a detection. This was done by subtracting the Bayesian log-evidence of a retrieval performed using minimal gas abundance from that resulting from a retrieval using the gas abundances listed earlier in this section. The resulting differences in log-evidences yields the log-Bayes Factor \citep[$lnB$;][]{2013ApJ...778..153B}. Log-Bayes Factors greater than 1 represent a weak detection, those greater than 2.5 a moderate detection, and those greater than 5 a strong detection \citep[see Table 2 of][]{2013ApJ...778..153B}. Bayes Factors are not meaningful for non-gaseous parameters because minimal values of these do not represent the absence of features, so this statistic was not calculated for $C_f$, $P_0$, $g$, or $A_s$.\par

\subsection{Optimal Bandpass Study Results} \label{bp results}
When examining the efficacy of these four instrumental designs, we are primarily concerned with maximizing the number of parameters constrained at the bandpass center with the shortest central wavelength possible. Minimizing the wavelength of the bandpass center is important for several reasons.  First, the throughput of a coronagraph is proportional to $\lambda / D$ due to the impact of the angular resolution of the telescope. Decreasing the wavelength ($\lambda$) improves the angular resolution of the telescope, allowing for planets closer to their hosts stars to be imaged. The habitable zones (HZ) of smaller stars or those further away are smaller in angular separation, so the ability to resolve planets closer to their host stars enables the characterization of HZ planets of smaller and/or more distant stars. Second, for G- and K-type stars, the stellar SED peaks at $0.5 - 0.7 \mathrm{\mu} m$, so positioning our bandpass closer to that wavelength region allows for a greater number of photons in reflected light, increasing the S/N achievable with the same exposure time.  

\begin{figure}[H]
\plotone{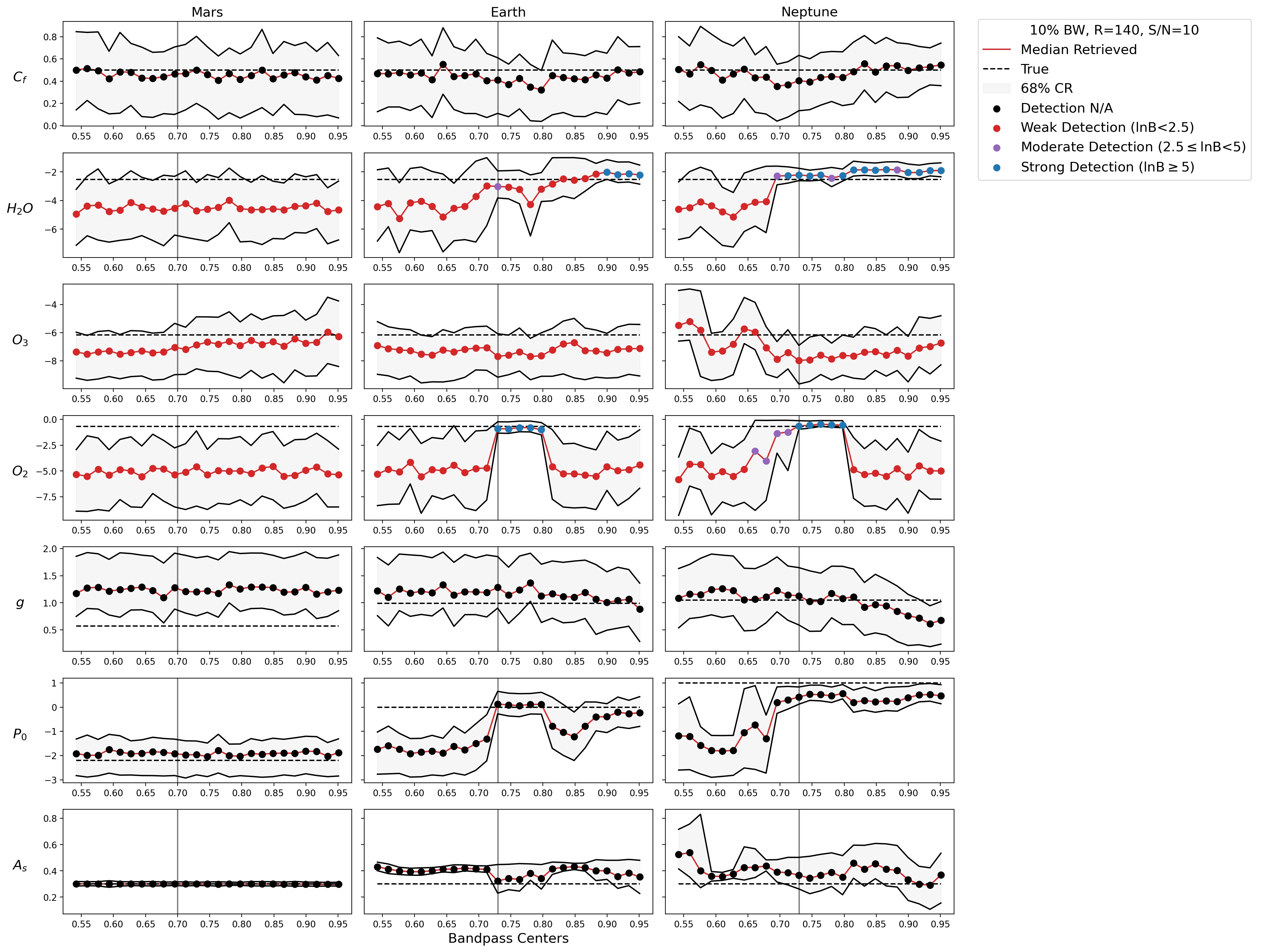}
\caption{Results of the bandpass study for the fiducial case. The grey regions represent the upper and lower limits of the 68\% credible region and the red line shows the (median) retrieved values with dots at each bandpass center. The true value for each parameter is shown as a horizontal dashed black line. Regions where the grey region narrows indicate the increased certainty of the Bayesian retrieval algorithm. Colored points along the red line indicate the strength of a detection of the particular parameter at each bandpass center. Red points indicate a weak detection ($lnB<2.5$), purple indicate a moderate detection ($2.5 \leq lnB < 5$), and blue indicates a strong detection ($lnB>5$). Black points, present for each non-gaseous parameter, represent the non-applicability of Bayes factors to these parameters. Grey vertical lines mark the minimum bandpass centers for each planet where the most parameters can be constrained.}
\label{fig:BF_bp_bw=0.1_R=140_SNR=10}
\end{figure}

We start by examining the fiducial case of the bandpass study, with a bandwidth of 10\%, S/N$=$10, and R$=$140. Figure~\ref{fig:BF_bp_bw=0.1_R=140_SNR=10} depicts these results. Under the Mars-like scenario, no constraints on any of the parameters are achieved except for $A_s$. This is likely due to the low surface pressure of this case, which diminishes the spectral features of the gases present in the atmosphere. This in turn makes $P_0$ difficult to constrain, which relies on the depth of the features. $A_s$ is constrained in every bandpass. In the Earth-like case, a strong detection of $O_2$ can be achieved using a bandpass centered at $0.73 \mathrm{\mu} m$, and simultaneously yield a moderate detection of $H_2O$. Alternatively, a strong detection of $H_2O$ can be made at $0.90 \mathrm{\mu} m$, but at the cost of a detection of $O_2$. Under the Neptune-like case, with its high surface pressure and therefore increased feature depth, constraints on $H_2O$ and $O_2$ can be made over a wider span of bandpass centers, but the minimum bandpass center which can be used to constrain both $H_2O$ and $O_2$ is still  $0.73 \mu m$. Constraints on surface pressure and albedo can also be made at similar bandpass centers for both the Earth- and Neptune-like cases.

\begin{figure}[H]
\plotone{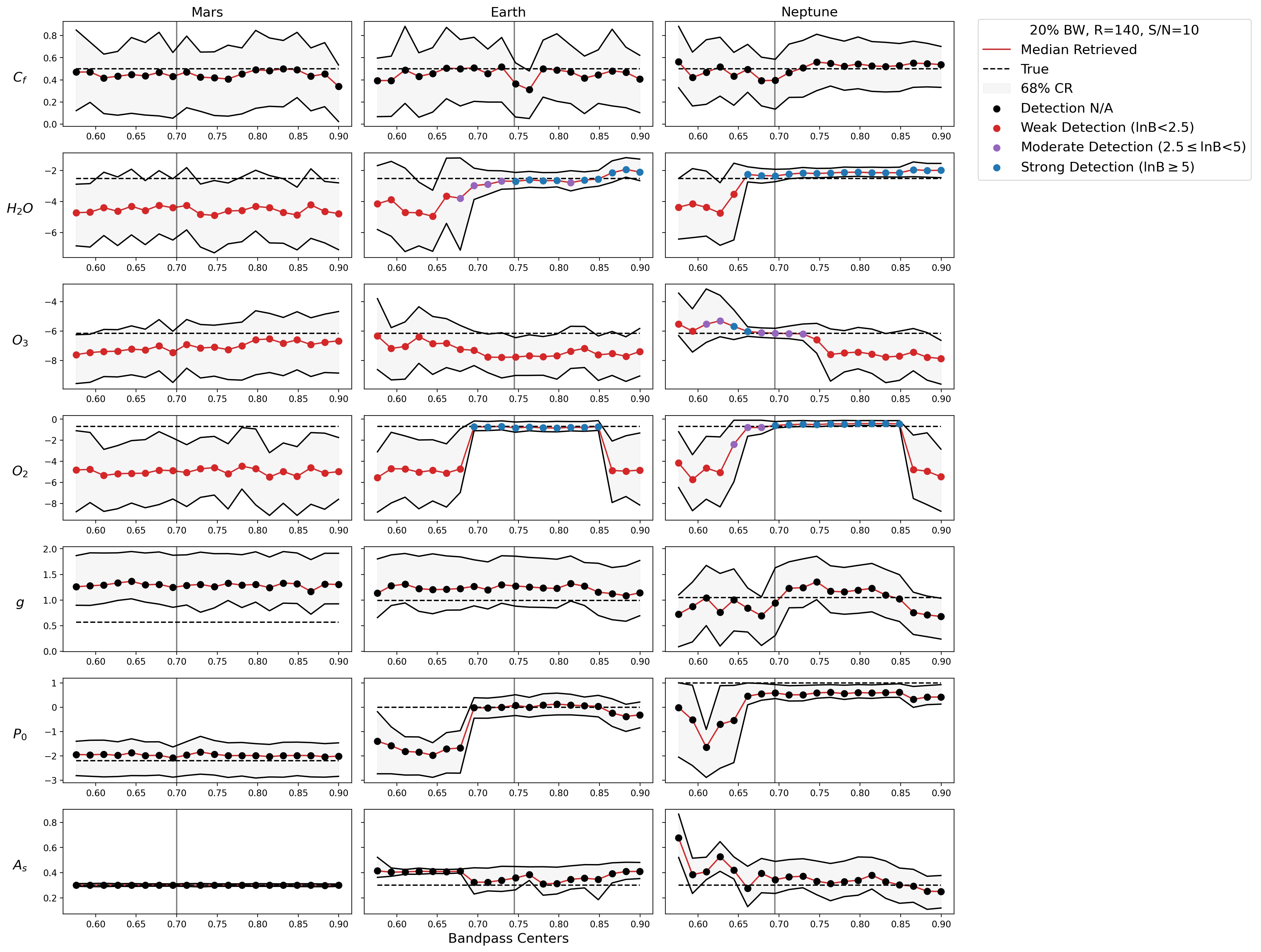}
\caption{Results of the bandpass study for the case with 20\% bandwidth. Aspects of this plot are explained in the caption below Figure~\ref{fig:BF_bp_bw=0.1_R=140_SNR=10}.}
\label{fig:BF_bp_bw=0.2_R=140_SNR=10}
\end{figure}

Next, we examine the case where the bandwidth is 20\% while S/N and R are the same as the fiducial values in Figure~\ref{fig:BF_bp_bw=0.2_R=140_SNR=10}. A wider bandwidth allows for a greater portion of a given spectral feature to be observed, thus yielding higher constraints on the gaseous parameters. While the Mars-like scenario still yields no constraints, the Earth-like scenario shows that strong constraints can be obtained for both $H_2O$ and $O_2$ using a bandpass centered at $0.74 \mu m$. The Neptune-like scenario shows strong detections of both $H_2O$ and $O_2$ as well as a moderate detection of $O_3$ when the bandpass is centered at $0.70 \mu m$. Detections of $P_O$ and $A_s$ are also likely in both the Earth and Neptune scenarios.

\begin{figure}[H]
\plotone{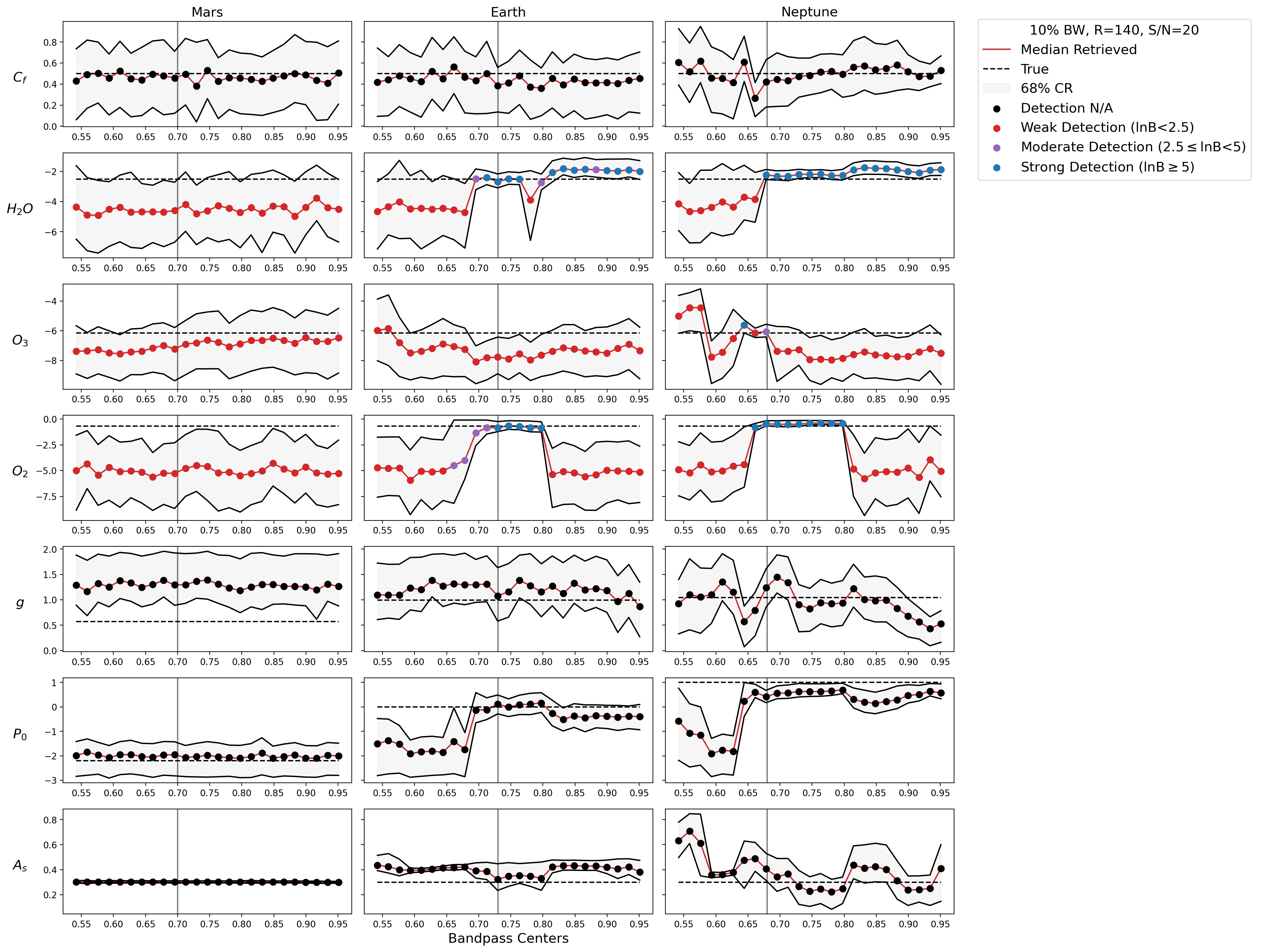}
\caption{Results of the bandpass study for the case with S/N=20. Aspects of this plot are explained in the caption below Figure~\ref{fig:BF_bp_bw=0.1_R=140_SNR=10}.}
\label{fig:BF_bp_bw=0.1_R=140_SNR=20}
\end{figure}

We also examine the case where the S/N is 20 while the bandwidth and R are the same as the fiducial values through Figure~\ref{fig:BF_bp_bw=0.1_R=140_SNR=20}. Increasing the S/N decreases the uncertainty in the retrieval and enables the sampling algorithm to obtain better constraints on the retrieved parameters. As before, the Mars-like case shows no constraints on any parameter except for $A_s$. In the Earth-like case, strong detections of $H_2O$ and $O_2$ can be made using a bandpass centered at $0.73 \mu m$. Additionally, strong detections of $H_2O$ can be made for bandpasses centered above $0.82 \mu m$. $H_2O$ and $O_2$ can be strongly detected on a Neptune-like planet using a bandpass centered at $0.68 \mu m$ and a moderate detection of $O_3$ can be made here as well. Surface pressure can likely be obtained from these same observations on the Earth- and Neptune-like planets, but a detection of surface albedo on these planets is less likely.

\begin{figure}[H]
\plotone{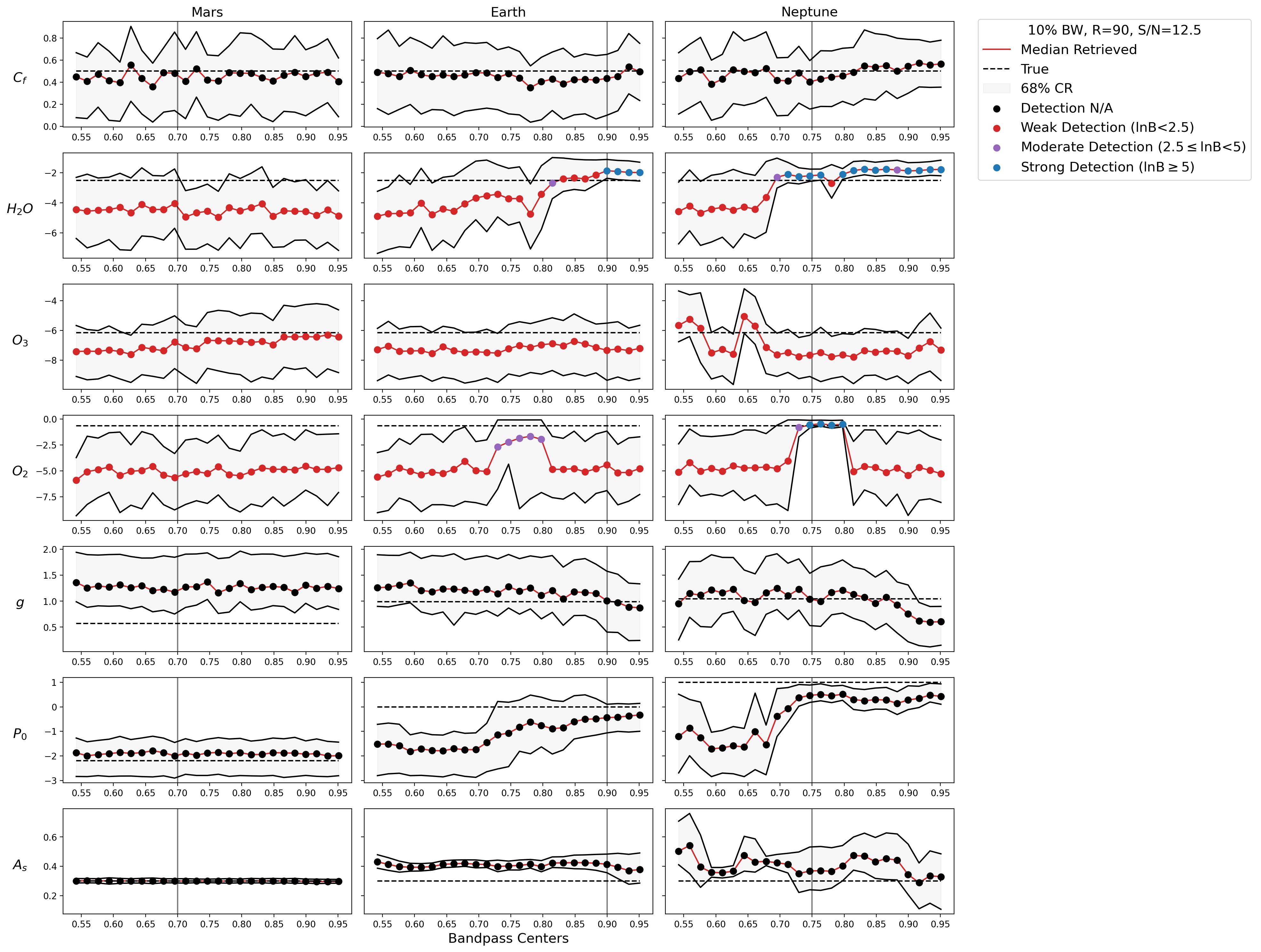}
\caption{Results of the bandpass study for the case with R=90 and S/N=12.5. Aspects of this plot are explained in the caption below Figure~\ref{fig:BF_bp_bw=0.1_R=140_SNR=10}.}
\label{fig:BF_bp_bw=0.1_R=90_SNR=12.5.jpg}
\end{figure}

Lastly, we examine the case where R=90 and the bandwidth is the same as the fiducial value in Figure~\ref{fig:BF_bp_bw=0.1_R=90_SNR=12.5.jpg}. Reducing the spectral resolution reduces the information content of the spectrum, causing fewer data points to be present within spectral features. This effect makes parameters more difficult to constrain using retrieval methods. Here, we adjust the S/N to be 12.5, in order to compare scenarios assuming a constant exposure time (under the assumption that the uncertainty is dominated by photon-noise statistics). Changing the S/N compensates for reduction in  resolving power from 140 to 90, so the photons collected in each bin increase by a factor of 140/90. Since we assume the observations are photon-noise-limited, this would increase the S/N by a factor of sqrt(140/90) = 1.25 for the same exposure time, meaning that S/N of 10 would become 12.5.  Like before, only $A_s$ is constrained in the Mars-like case. The Earth-like case yields a strong detection of $H_2O$ when the bandpass is centered on $0.90 \mu m$, but no other strong detections are available. The Neptune-like case can obtain strong detections of $H_2O$ and $O_2$ using a bandpass centered at $0.75 \mu m$, while strong detections of $H_2O$ are still possible using bandpasses centered a wavelengths higher than $0.80 \mu m$. $A_s$ is potentially well-constrained in the Earth-like case while $P_0$ can likely be constrained in the Neptune-like case.

To take a closer look at how the posteriors of certain parameters are affected by the bandpass center, width, R, and S/N, we plotted the marginalized posteriors of $H_2O$, $O_3$, $O_2$, and $P_0$ from the test cases atop those from the fiducial case. These posteriors are the results of the retrievals performed at the bandpass centers positioned at the vertical lines in Figures~\ref{fig:BF_bp_bw=0.1_R=140_SNR=10} through~\ref{fig:BF_bp_bw=0.1_R=90_SNR=12.5.jpg}. We only show the posteriors of these parameters for the Earth-like and Neptune-like planets because the no parameters were well-constrained in any of the retrievals on the Mars-like planet (except $A_s$). These plots show the best-case (in terms of minimal bandpass center) retrievals for the particular bandwidth, R, and S/N cases.

\begin{figure}[H]
\plotone{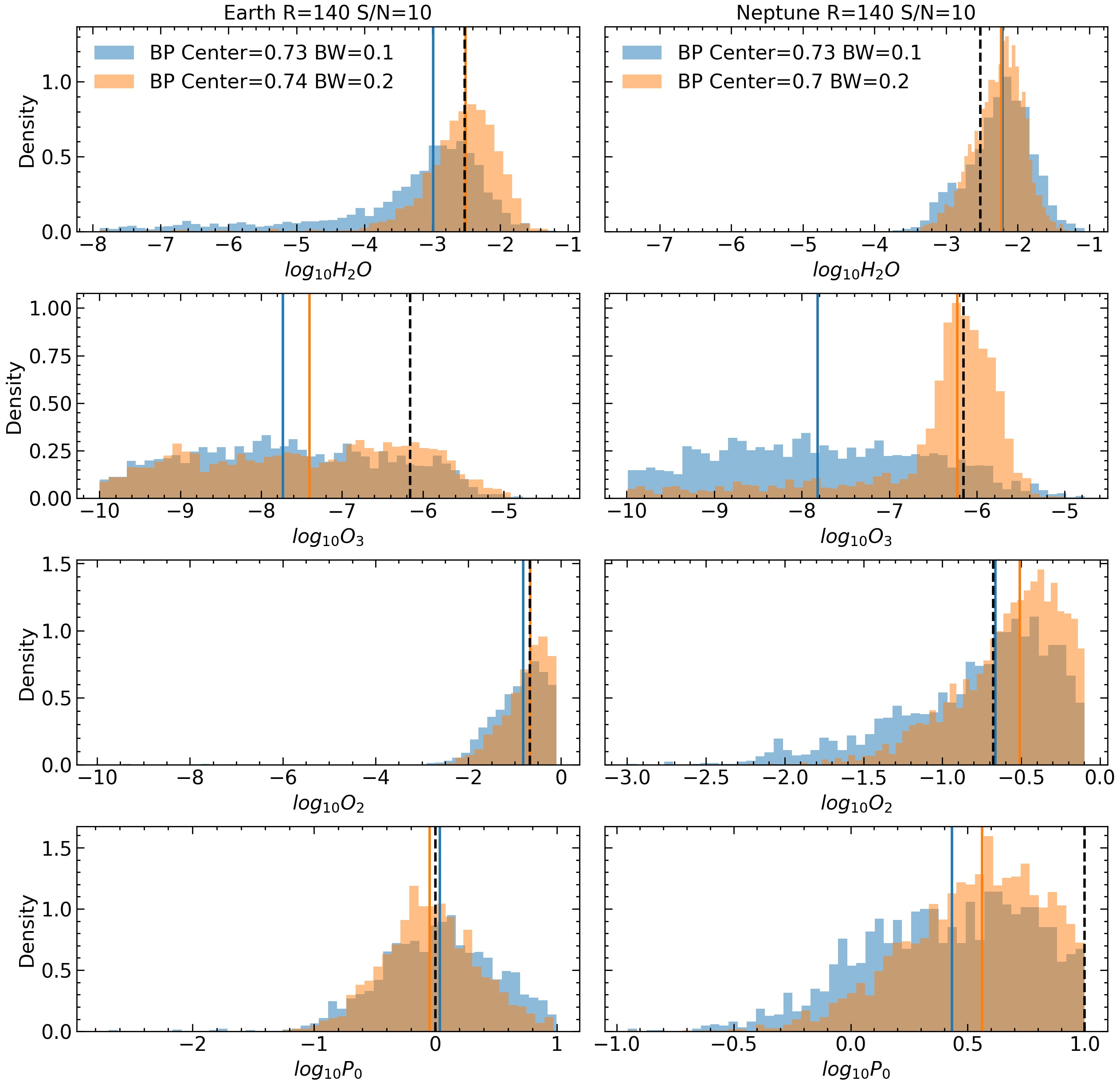}
\caption{Posteriors of the bandpass studying comparing the fiducial to the 20\% bandwidth case. The marginalized posteriors for $H_2O$, $O_3$, $O_2$, and $P_0$ are shown in blue for the fiducial (R=140, S/N=10, BW=0.1) case and orange for the 20\% bandwidth case. Blue and orange vertical lines show the median values of these retrieval results while a dashed black line shows the true value.}
\label{fig:post_bw}
\end{figure}

First, Figure~\ref{fig:post_bw} compares the posteriors of the fiducial case to those of the 20\% bandwidth case. Increasing the bandwidth causes more spectral features to be included in a particular wavelength range, so parameters with narrow features will be better constrained. $H_2O$, $O_2$, and $P_0$ were shown to be well constrained in Figures~\ref{fig:BF_bp_bw=0.1_R=140_SNR=10} and~\ref{fig:BF_bp_bw=0.2_R=140_SNR=10} and these posteriors are consistent with this finding. The constrains represented in the Neptune-like scenario of Figure~\ref{fig:BF_bp_bw=0.2_R=140_SNR=10} are represented here as well. The $O_3$ panel of the Neptune-like column (second panel from the top in the second column) shows that $O_3$ is well constrained using a bandwidth of 20\% centered at $0.70 \mu m$ while using a bandwidth half as wide with a bandpass centered at $0.73 \mu m$ yields no $O_3$ constraint.

\begin{figure}[H]
\plotone{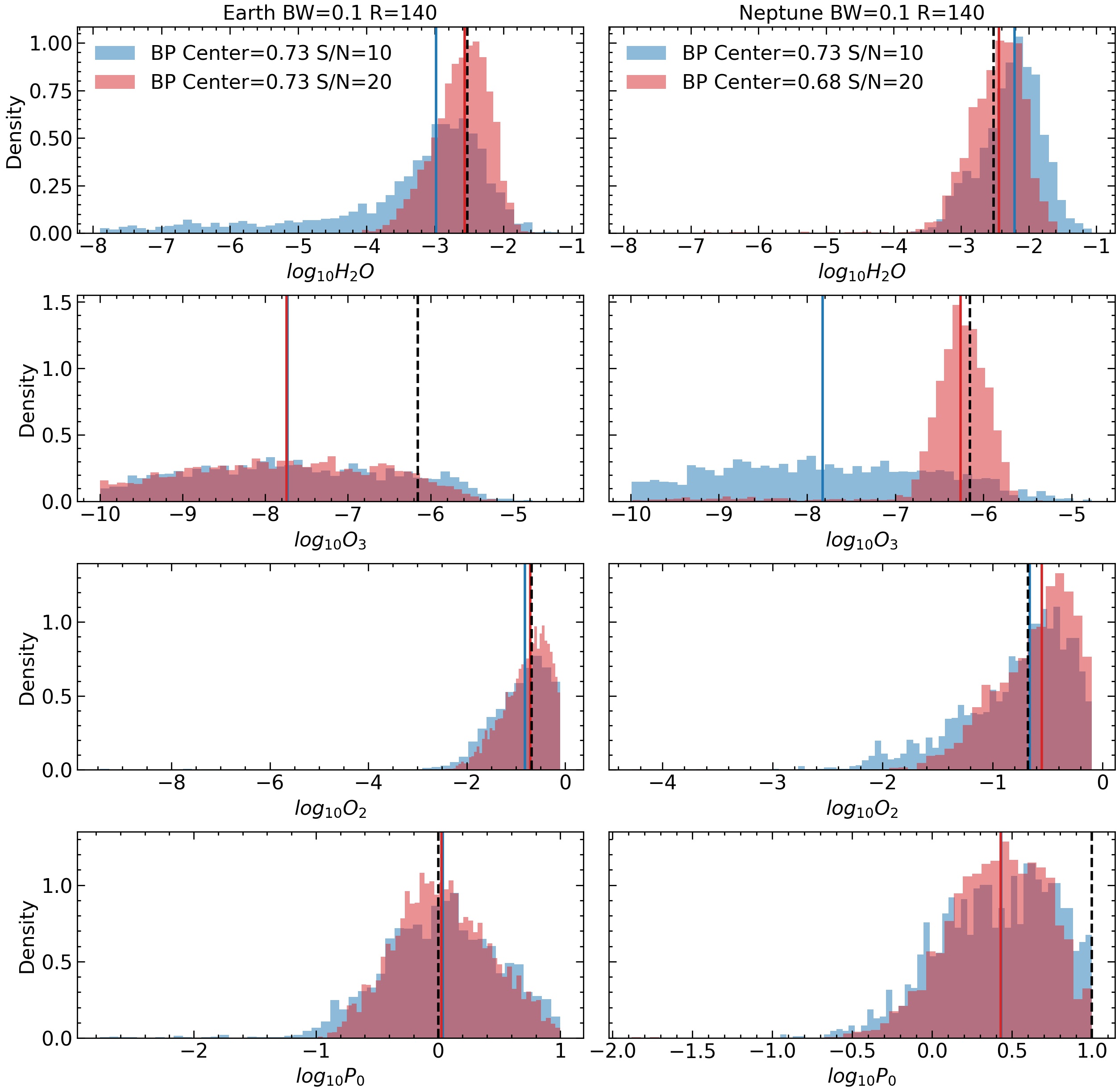}
\caption{Posteriors of the bandpass studying comparing the fiducial to the S/N=20 case. The marginalized posteriors for $H_2O$, $O_3$, $O_2$, and $P_0$ are shown in blue for the fiducial (R=140, S/N=10, BW=0.1) case and red for the 20\% bandwidth case. Blue and red vertical lines show the median values of these retrieval results while a dashed black line shows the true value.}
\label{fig:post_SNR}
\end{figure}

Following this, we compare the posteriors of the fiducial case to those of the S/N=20 case using Figure~\ref{fig:post_SNR}. Increasing the S/N decreases the uncertainty of the retrieval, so the retrieved (median) values are closer to their respective true values in most cases. The benefit of increasing the S/N is clear in the $O_3$ panel of the Neptune-like column (second panel from the top in the left column). Increasing this instrumental parameter leads to a posterior which is much more tightly confined around the true value.

\begin{figure}[H]
\plotone{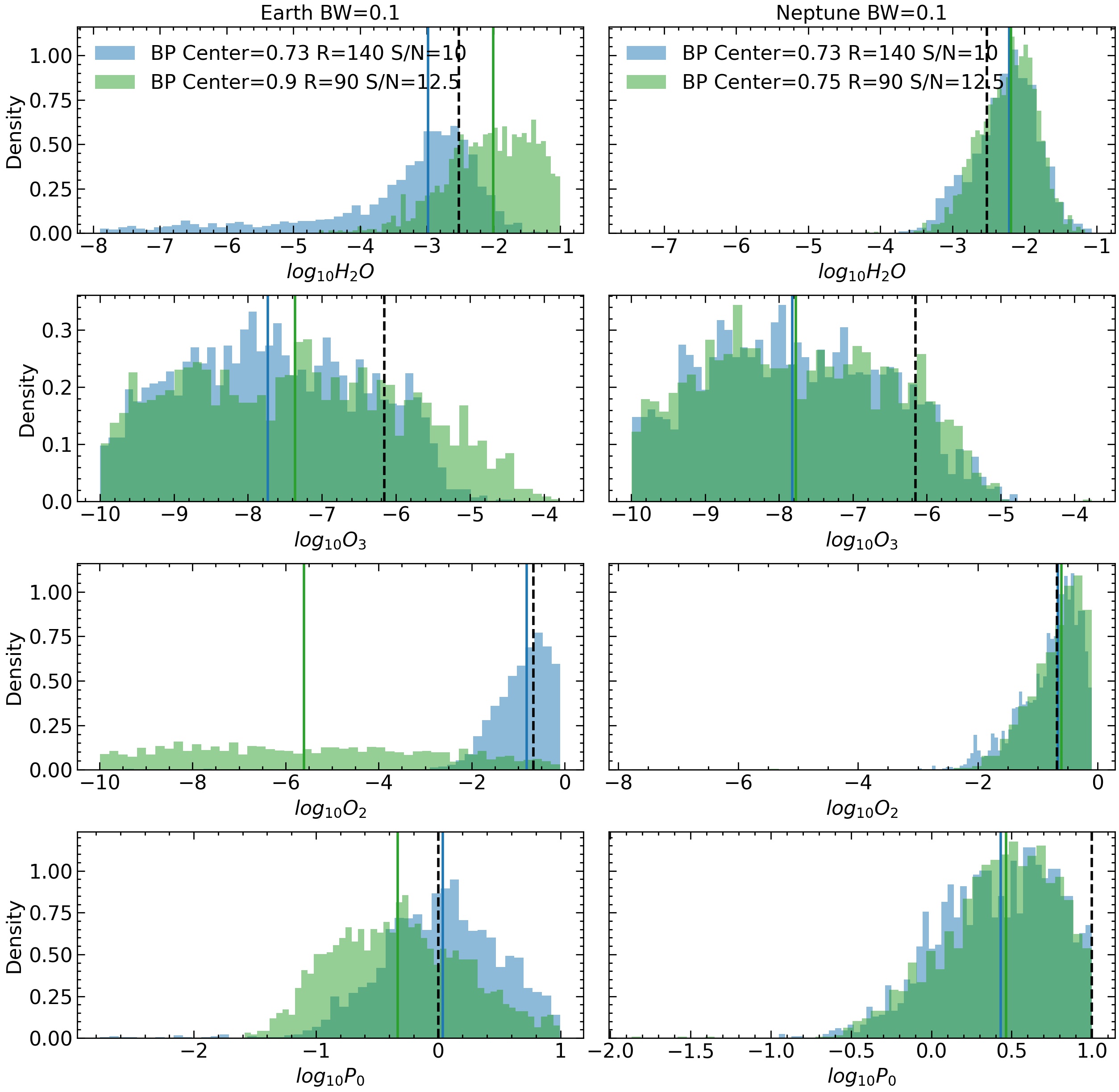}
\caption{Posteriors of the bandpass studying comparing the fiducial to the R=90 and S/N=12.5 case. The marginalized posteriors for $H_2O$, $O_3$, $O_2$, and $P_0$ are shown in blue for the fiducial (R=140, S/N=10, BW=0.1) case and green for the 20\% bandwidth case. Blue and green vertical lines show the median values of these retrieval results while a dashed black line shows the true value.}
\label{fig:post_R}
\end{figure}

Lastly, we compare the fiducial posteriors to those resulting from the R=90 case in Figure~\ref{fig:post_R}. Decreasing the resolving power decreases the information content of the spectrum, making the retrieval of narrow spectral features more difficult. This is true in the Earth-like case where $O_2$ can no longer be constrained when the resolution is decreased. The $O_2$ panel of the Earth-like case (third panel from the top in the right column) shows a totally unconstrained posterior for the R=90 case while the fiducial case (with R=140) is exists tightly around the true value.

These results are summarized in Table~\ref{tab:BF_bp_results_mars} for the Mars-like scenario, Table~\ref{tab:BF_bp_results_earth} for the Earth-like scenario, and Table~\ref{tab:BF_bp_results_nept} for the Neptune-like scenario.

\begin{deluxetable}{ccc} \label{tab:bp_results_mars}
\tablecaption{Bandpass Study Results (Mars-like Scenario)}
\label{tab:BF_bp_results_mars}
\tablewidth{0pt}
\tablehead{
\colhead{Case} & \colhead{Params. Constrained} & \colhead{Min. Bandpass Center ($\mu$m)}
}
\startdata
10\% BW, R=140, S/N=10 & None & N/A \\
20\% BW, R=140, S/N=10 & None & N/A \\
10\% BW, R=140, S/N=20 & None & N/A \\
10\% BW, R=90, S/N=12.5 & None & N/A \\
\enddata
\tablecomments{Only strong detections ($lnB>5$) are shown.}
\end{deluxetable}

\begin{deluxetable}{ccc} \label{tab:bp_results_earth}
\tablecaption{Bandpass Study Results (Earth-like Scenario)}
\label{tab:BF_bp_results_earth}
\tablewidth{0pt}
\tablehead{
\colhead{Case} & \colhead{Params. Constrained} & \colhead{Min. Bandpass Center ($\mu$m)}
}
\startdata
10\% BW, R=140, S/N=10 & $O_2$ & 0.73 \\
20\% BW, R=140, S/N=10 & $H_2O$, $O_2$ & 0.74 \\
10\% BW, R=140, S/N=20 & $H_2O$, $O_2$ & 0.73 \\
10\% BW, R=90, S/N=12.5 & $H_2O$ & 0.90 \\
\enddata
\tablecomments{Only strong detections ($lnB>5$) are shown.}
\end{deluxetable}

\begin{deluxetable}{ccc} \label{tab:bp_results_nept}
\tablecaption{Bandpass Study Results (Neptune-like Scenario)}
\label{tab:BF_bp_results_nept}
\tablewidth{0pt}
\tablehead{
\colhead{Case} & \colhead{Params. Constrained} & \colhead{Min. Bandpass Center ($\mu$m)}
}
\startdata
10\% BW, R=140, S/N=10 & $H_2O$, $O_2$ & 0.73 \\
20\% BW, R=140, S/N=10 & $H_2O$, $O_2$ & 0.70 \\
10\% BW, R=140, S/N=20 & $H_2O$, $O_2$ & 0.68 \\
10\% BW, R=90, S/N=12.5 & $H_2O$, $O_2$ & 0.75 \\
\enddata
\tablecomments{Only strong detections ($lnB>5$) are shown.}
\end{deluxetable}

At the current stage in the process of developing the next generation of space telescopes, instrumental designs have not yet been determined. What we know for certain is that observing time will be a limited resource. Therefore, we use the results of this bandpass study to recommend instrumental designs (in terms of bandwidth and R) that will yield the most molecular detections while minimizing observation time (which is proportional to S/N). We focus these recommendations on the Earth-like scenario and prioritize observations centered on shorter wavelengths. Increasing the exposure time of an observation such that the S/N becomes 20 allows for constraints on both $H_2O$ and $O_2$ for observations using a resolution of 140 and bandwidth of 10\% centered at 0.73 $\mu m$, but these same molecular detections can also be made by changing instrumental parameters other than exposure time. If a bandwidth of 20\% is used alongside a resolution of 140, then both $H_2O$ and $O_2$ can be detected when the S/N is only 10 for observations are centered at 0.73 $\mu m$. If a bandwidth of 10\% is used, then a resolution greater than 90 and less than or equal to 140 is needed to detect both molecules. If the bandwidth and resolution are set to 10\% and 90 respectively, then only $H_2O$ can be detected and at a higher wavelength of around 0.90 $\mu m$. Increasing the bandwidth of future instruments may be the best way to detect a greater number of molecules, while lowering the resolution or bandwidth will necessitate a greater exposure time to achieve the same detections. If only a detection of $O_2$ is required, then this can be achieved using observations centered at wavelengths as short as 0.69 $\mu m$ if the bandwidth is 20\%. Additional work is needed to determine feasible levels resolution and bandwidth parameter can varied. Future works should also increase the fidelity of these experiments.

\section{Discussion} \label{discussion}

\subsection{Simplifying Assumptions} \label{discussion3}
Several simplifications were employed to reduce the computational complexity of our simulated spectra. Firstly, each simulated cloudy atmospheres uses an isotropic distribution of clouds. We considered using distinct cloud layers but this was causing issues at higher surface pressures, and it would add a high level of model dependence on the microphysical assumptions regarding these clouds. The assumed noise in the retrievals performed throughout this work is constant and wavelength-independent, which allow us to better separate modeling/grid errors from a-priori noise considerations.

\subsection{Grid Construction Methods} \label{discussion1}
One of the greatest challenges we encountered while building the grid was devising an error metric that would accurately reflect the degree of the interpolation error present in intermediate spectra during retrieval. Retrievals cannot be performed using the grid until the grid is made, yet assessing the performance of the grid in this context is challenging until it can be used in the retrievals themselves. We approached this by calculating interpolation errors at different levels of off-axis parameters, but this does not truly account for the degeneracies present in the retrieval when all parameters are free to vary. Additionally, the choice of an error metric is one that could be further optimized. We attempted to devise a metric which captures the most significant deviations from the true spectrum relative to the true spectrum itself, but found that this error metric we used when building the grid does not correlate well with the interpolation error found when evaluating the grid after it was built. Instead, an error metric which resembles the log-likelihood function (e.g. sum of square errors) may be more closely related to retrieval performance. Furthermore, the error cutoff level which is used to terminate the grid construction algorithm could also be altered. We chose to terminate the construction process at when the top deviations are 10\% of the ground truth, but the contrived nature of our error metric limits the interpretability of this cut-off. Future efforts may further explore alternative metrics which are able to better predict error in retrievals before the grid is constructed.


\subsection{Grid Based Retrievals as a Means to Investigate Multiple Noise Realizations} \label{scatter}
Other works \citep[e.g.,][]{2018AJ....155..200F} have discussed the importance of adding randomized noise to the data spectrum and running multiple retrievals with different noise realization. While this method may be an ideal way to simulate more realistic observations, it is often not done due to the long runtimes associated with running multiple retrievals. However, the grid-based methods discussed here are able to easily perform multiple retrievals using different noise realizations because of the remarkably quick runtime this method enjoys (on the order of seconds). Despite having the capability to enhance the fidelity of our retrievals in this way, we chose to perform retrievals on data spectra without any additional scatter added to the spectral data points. This decision was made to better isolate the numerical effects of multi-dimensional interpolation. Introducing random noise to the relatively small number of spectral points in each bandpass could result in any unusually large stochastic fluctuations in the artificial scatter. This, or any inherent bias in the random number generation, could induce systemic effects on the results of the retrievals which would be difficult to identify and account for across all of the investigations conducted for this study. Future work could use our grid-based methods to study the effect that multiple realizations of random noise would have on retrieval results.

\subsection{Grid Point Placement} \label{discussion4}
While we adopted an iterative approach which constructs a grid by placing points at the location of highest interpolation error, other works have proposed different methods. \cite{https://doi.org/10.48550/arxiv.2206.12194} investigated grid point placement by random and Latin hypercube (LH) sampling and compared these methods to traditional evenly-spaced linear sampling \citep{Allard_2001, 10.1093/mnras/sty3001, 2021ApJ...920...85M}. They found that grids produced using random or LH sampling outperform those produced using linear spacing for all but the lowest grid dimensionalities. Unfortunately, direct comparisons between our results and those of \cite{https://doi.org/10.48550/arxiv.2206.12194} are difficult for multiple reasons. First, they compute spectra intermediate to grid points using a random forest machine learning model whereas we used linear interpolations. These techniques may perform fundamentally differently in such a way that one form of grid sampling is optimal for one method while another form of grid sampling is optimal for the other - future work should investigate this. Second, their grids implement different atmospheric parameters than ours. As shown in Figure~\ref{fig:interp err}, the interpolation error can vary greatly between parameters. While they found random and LH grid point sampling to perform better than linear sampling (especially at higher grid dimensions), this approach should be used with caution when considering parameters with highly nonlinear effects such as $O_3$. Random sampling could easily miss critical points between spectra morphologies caused by subtle differences in parameters like $O_3$. However, random and LH sampling are advantageous for grids which include a large number of parameters as our iterative sampling approach (as well as linear methods) is susceptible to the curse of dimensionality.

\section{Conclusions} \label{conclusions}
Throughout this work, we have demonstrated methods for constructing, validating, and deploying precomputed grids of model spectra for use in the atmospheric retrievals of exoplanets. Interpolating spectra from a grid and using these as the models within a Bayesian framework significantly accelerates retrievals compared to traditional methods of calculating each model spectrum on-demand, reducing runtimes from days or weeks to seconds or minutes using a standard high-end laptop. Though interpolation from a grid will induce some error into the model comparison spectrum and will therefore prove to be somewhat less accurate than a true radiative transfer calculation, the extreme efficiency of grid-based retrievals can enable a host of new studies that were previously thought to be computationally infeasible. The methods presented here can be used to make any other grid with any other combination of variables, in which case some of the same issues we observed in this work may no longer be relevant (Figure ~\ref{region}).

Our evaluation procedure reveals that the linear interpolation between spectral grid points for our grid are generally very accurate (average $R^2$=0.998) and retrieval performance is not significantly inhibited by this interpolation error (Figure~\ref{fig:retr_eval}). However, there are particular regions within the parameter space explored in this work which are problematic for retrievals. This is due in part to the interpolation error of these regions (Figure~\ref{fig:interp err} and also the degenerate nature of spectra in this region (Figure~\ref{fig:O3_spectra}. Future works should avoid using this grid for retrievals involving concurrently high $O_3$ ($log_{10}O_3>-5$), low $P_0$ ($log_{10}P_0<-1.7$), and high $A_s$ ($A_s=0.96$). Furthermore, we have shown the grid-based techniques can be used to enable a variety of studies which were previously regarded as computationally infeasible, such as simulating yields from future observations using a variety of instrumental setups. Our techniques for constructing and evaluating model grids can be applied to a wide variety of use cases. For example, they could be used to improve the grid-based methodologies employed in the James Webb Space Telescope Early Release Science spectral analysis by enabling the construction of chemistry model grids with minimal interpolation error and thereby improve the accuracy of their retrievals \citep{2022arXiv221110488A, 2022arXiv221110487R}.  Future works may choose to use the grid presented in this work or employ our methods to build a grid of their own. \par

As a first application of the capabilities of our grid-based retrieval methods using our 6-parameter grid, we conducted an examination of the sensitivity of retrieval results to instrument and observation design parameters. Many details pertaining to the design of future observatory instruments have not yet been determined, and we can utilize a yield analysis of different instrument configurations in order to help inform these decisions. We performed a sequence of retrievals using simulated observations of Solar System planet analogues centered at 25 different wavelength positions within the proposed LUVOIR visible-light wavelength range of 0.515-1.0 $\mu m$. Four instrumental designs were explored which varied the spectral bandwidth, resolution, and exposure time (via S/N). We found that detections of $H_2O$ and $O_2$ in the atmospheres of Earth-like planets can be made using observations centered at 0.74 $\mu m$ simultaneously if a 20\% bandwidth is used. Using an S/N of 20 can yield detections of the same molecules at a similar bandpass center. If the resolution is 90, then only $H_2O$ can be detected and at a much longer wavelength of $0.90 \mu m$. From these results, we conclude that detections of both $H_2O$ and $O_2$ are obtainable on Earth-like planets if only the bandwidth is increased; increasing exposure time is not necessary. Similar tests performed using resolutions between 90 and 140 are needed to determine the minimum resolution necessary to detect both $H_2O$ and $O_2$ when the bandwidth is 20\%. In general, a broader variety of instrument parameters should be examined using methods like those shown here in order to make more definitive recommendations for future designs. Future work could improve on this by increasing the fidelity of these experiments.

The authors would like to thank the Sellers Exoplanet Environments Collaboration (SEEC) and ExoSpec teams at NASA's Goddard Space Flight Center for their consistent support.


\bibliography{main}{}
\bibliographystyle{aasjournal}

\end{document}